\documentclass[journal]{IEEEtran}
\usepackage{amsmath,amsfonts,amssymb, dsfont}

\usepackage{tabularx}
\usepackage{multirow} 
\usepackage{booktabs} 
\usepackage{diagbox} 

\usepackage{cite}
\usepackage{hyperref}

\usepackage{graphicx}
\makeatletter
    \let\MYcaption\@makecaption
\makeatother
    \usepackage[font=footnotesize]{subcaption}
\makeatletter
    \let\@makecaption\MYcaption
\makeatother

\usepackage{tikz, pgfplots}
\usetikzlibrary{shapes, arrows, arrows.meta, fit, positioning, calc, 3d}
\usepgfplotslibrary{fillbetween}
\pgfplotsset{compat=1.18} 
\tikzset{>=latex}

\DeclareMathOperator*{\argmin}{arg\,min}

\newcommand{\set}[1]{\mathcal{#1}}

\allowdisplaybreaks

\begin{document}
\bstctlcite{IEEEexample:BSTcontrol}

\title{From Information Freshness to Semantics of Information and Goal-oriented Communications}
\author{Jiping~Luo, Erfan~Delfani, Mehrdad~Salimnejad, and Nikolaos~Pappas
\thanks{The authors are with the Department of Computer and Information Science, Link\"oping University, Link\"oping 58183, Sweden (e-mail: jiping.luo@liu.se; erfan.delfani@liu.se; mehrdad.salimnejad@liu.se; nikolaos.pappas@liu.se). This work has been supported by the Swedish Research Council (VR), ELLIIT, the European Union (ETHER, 101096526, 6G-LEADER, 101192080, ELIXIRION, 101120135, and ROBUST-6G, 101139068), and the Graduate School in Computer Science (CUGS).}
}
\IEEEspecialpapernotice{(Invited Paper)}

\maketitle
\begin{abstract}
Future wireless networks must support real-time, data-driven cyber-physical systems in which communication is tightly coupled with sensing, inference, control, and decision-making. Traditional communication paradigms centered on accuracy, throughput, and latency are increasingly inadequate for these systems, where the value of information depends on its semantic relevance to a specific task. This paper provides a unified exposition of the progression from classical distortion-based frameworks, through information freshness metrics such as the Age of Information (AoI) and its variants, to the emerging paradigm of goal-oriented semantics-aware communication. We organize and systematize existing semantics-aware metrics, including content- and version-aware measures, context-dependent distortion formulations, and history-dependent error persistence metrics that capture lasting impact and urgency. Within this framework, we highlight how these metrics address the limitations of purely accuracy- or freshness-centric designs, and how they collectively enable the selective generation and transmission of only task-relevant information. We further review analytical tools based on Markov decision process (MDP) and Lyapunov optimization methods that have been employed to characterize optimal or near-optimal timing and scheduling policies under semantic performance criteria and communication constraints. By synthesizing these developments into a coherent framework, the paper clarifies the design principles underlying goal-oriented, semantics-aware communication systems. It illustrates how they can significantly improve efficiency, reliability, and task performance. The presented perspective aims to serve as a bridge between information-theoretic, control-theoretic, and networking viewpoints, and to guide the design of semantic communication architectures for 6G and beyond.
\end{abstract}

\begin{figure*}[t!]
    \centering
    \includegraphics[width=\linewidth]{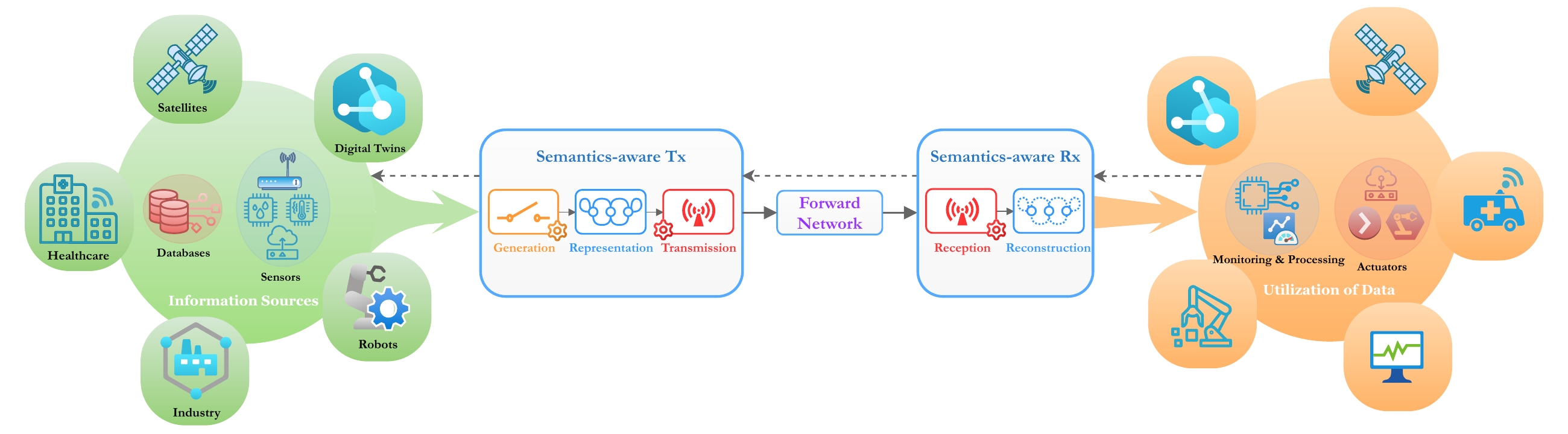}
    \caption{End-to-end semantics-aware goal-oriented communication system.}
    \label{fig_HighLevelSystem}
\end{figure*}

\section{Introduction}
The surge in mobile traffic, the pursuit of ubiquitous connectivity, and the rapid expansion of data generation are natural consequences of the ongoing digital information revolution. These trends will intensify in the upcoming era of connected intelligence, where networks of autonomous devices such as robots, vehicles, drones, and other cyber–physical agents operate with advanced sensing, communication, computing, and learning capabilities~\cite{vitturi2013industrial, park2018wireless, gielis2022critical}. As a result, future wireless systems must efficiently manage vast, distributed, and heterogeneous data streams while supporting real-time operation to \emph{deliver timely information} and \emph{promptly respond} to critical events~\cite{gielis2022critical, kountouris2021semantics, Popovski2020Indian}. In such environments, communication is no longer a passive data pipe of reliable information delivery, but a decisive component of system functionality.

Uncoordinated or indiscriminate generation, storage, and transmission of massive data volumes for remote processing inevitably lead to network congestion and excessive energy consumption, thereby jeopardizing the stringent delay, reliability, and responsiveness requirements of emerging applications. Traditional communications systems are designed adopting a emph{maximalistic approach}, which exposes a fundamental inefficiency. Systems are engineered with ample performance margins, aggressive reliability targets, and extensive resource allocation, often leading to over-provisioning and limited scalability.

In contrast, \emph{goal-oriented semantic communication} advocates a \textit{minimalist design}, grounded in the principle that \textit{``less is more''} \cite{kountouris2021semantics, AgheliTCOM24}. Rather than maximizing data throughput or blindly ensuring high-fidelity reproduction, semantic communication focuses on transmitting only information that is \emph{useful} for accomplishing a given task. This perspective promises substantial gains in resource utilization, energy efficiency, and computational effectiveness. The vision can be traced to Weaver’s early commentary on Shannon’s foundational theory. Despite multiple historical attempts to enrich the conceptual framework of semantic communication, most remained abstract and lacked an operational formulation \cite{utkovski2023semantic}. Today, however, the emergence of connected intelligence, real-time cyber–physical systems, and interactive multi-agent networks has renewed the urgency for a practical and rigorous semantic communication theory.

Realizing this vision requires overcoming two fundamental challenges. The first is the integration of the real-time processes involved in generating, processing, transferring, and reconstructing multi-modal information, each of which affects the utility of the communicated data. The second is \emph{ensuring that the entire communication chain, from data inception to its final utilization, incorporates the semantic significance and utility of information}. Hence, it is no longer sufficient to view communication as a modular subsystem; rather, it must be treated as an active decision-making entity that orchestrates the information that flows through the network in an end-to-end manner.

These considerations motivate a deeper investigation into \emph{when, what, and how} data should be generated, processed, transmitted, and utilized, and how the effective value of information for a given task can be assessed. A core objective is to jointly optimize information generation (sensing), communication, and actuation decisions through the lens of semantic relevance. Early implementations of importance-aware, end-to-end communication systems demonstrate the potential of this paradigm by significantly reducing uninformative transmissions while maintaining, and often improving, system performance.

Despite the progress, existing state-of-the-art methodologies frequently overlook message content and its operational impact. While foundational steps have been taken, a universally accepted semantic communication theory remains elusive. Classical metrics such as throughput, packet loss, or delay are inherently semantics-agnostic; they quantify network performance but do not reflect how information contributes to decision-making, control actions, or collective intelligence. For applications involving real-time control, event detection, or distributed learning, the value of information is fundamentally tied to its correctness, context, and timeliness.

Metrics such as the \emph{Age of Information} (AoI) have emerged to capture information freshness, a critical attribute for many monitoring and inference tasks. However, \textbf{networks supporting heterogeneous, task-specific semantic requirements demand more advanced metrics that integrate freshness, content significance, contextual relevance, and task utility into comprehensive semantic evaluations}, along with access and scheduling mechanisms tailored to these enriched performance metrics. Developing such a framework is key to enabling the next generation of communication systems that are goal-oriented, in which the purpose of communication, rather than the quantity of transmitted data, shapes network design.

\section{Literature Overview}

\subsection{Information Accuracy} 
Information accuracy has long been the primary concern in the design of modern communication and control systems. In this paradigm, the central purpose is to ensure that the information delivered is adequately faithful to enable reliable inference, decision-making, and control~\cite{shannon, gallager1968information, Brockett-TAC-1997, schenato2007foundations, hespanha2007survey}. We briefly review the problem of real-time remote estimation and control of Markov processes, a canonical setting in many networked control systems (NCSs) and a motivating example for the discussions in this paper.

This line of research can be traced back to the design of real-time source coders~\cite{witsenhausen1979structure, walrand1983optimal}, in which the source outputs are causally encoded (compressed) and then sent to a receiver through a noisy channel. Coding and decoding operate in real time, and the distortion measure does not tolerate delays. More results on optimal real-time source coders can be found in, e.g.,~\cite{teneketzis2006structure, matloub2006universal, akyol2014zero, wood2017optimal}. In the control literature, several studies have developed optimal communication and estimation policies for Markovian systems. For example, remote estimation of a scalar Gaussian source with communication costs was examined in~\cite{lipsa2011remote}, where the authors proved that a threshold communication policy (i.e., the sensor transmits whenever the estimation error exceeds a threshold), together with a Kalman-like estimator, is jointly optimal. These results were further extended to systems with multidimensional Gaussian sources and energy-harvesting sensors~\cite{nayyar2013optimal}, communication resource constraints~\cite{chakravorty2017fundamental}, unreliable channels with adaptive noise~\cite{gao2018optimal}, and packet drops~\cite{chakravorty2020remote}, among others. 

Another line of work investigates the use of Kalman filters for remote estimation. In~\cite{sinopoli2004kalman}, the authors characterized the performance of Kalman filtering at a remote receiver under intermittent measurements. The design of optimal communication policies can be greatly simplified when the sensor runs a Kalman filter and sends its estimated states, rather than raw measurements, to the receiver; see, e.g.,~\cite{schenato2008optimal, wu2013event, leong2017sensor, wu2020learning, liu2022remote}. There is also a rich body of research on the estimation of discrete-state Markov chains (see, e.g.,~\cite{gauvain1994maximum, VVV-TIT-2010, Vikram-TIT-2013, chen2017event, chakravorty2020remote, liu2023toward}). Markov chains are often used in robust estimation and control systems to model disturbances and anomalies~\cite{costa2005mjls}.

This classical framework assumes that \emph{all information is equally important} and that the cost of an error depends solely on its physical discrepancy. However, this assumption warrants re-examination in emerging cyber-physical systems, where uniformly high fidelity across all information components may be wasteful, if not infeasible. This limitation has motivated research on unequal error protection (UEP)~\cite{masnick1967linear, borade2009unequal}, in which coding schemes allocate greater protection to information components of higher significance.

\subsection{Information Freshness} 
Timeliness is emerging as a crucial requirement in many cyber-physical systems such as intelligent transportation, industrial automation, and swarm robotics~\cite{jiping2023TITS, vitturi2013industrial, gielis2022critical, pappas2023age}. It is important to distinguish between timely and low-latency communications. The latter places a constraint on the \emph{packet-level} delay of a given data stream, whereas timeliness is a \emph{source-level} requirement concerning how fresh our knowledge is about an information source. Importantly, timely communication requires direct control over the generation of information~\cite{kaul2012real}.

The notion of AoI was introduced in~\cite{song1990performance, kaul2011minimizing, kaul2012real} to measure the freshness of information that a remote receiver has about an information source. The underlying assumption is that \emph{information holds the greatest value when it is fresh}. The work in~\cite{kaul2012real} initiated the study of AoI from a queueing-theoretic perspective. The average age, peak age, age distributions, and nonlinear age functions have been analyzed in various queueing systems with multiple servers, multiple hops, and/or multiple sources (see, e.g.,~\cite{kam2015effect, pappas2015age, huang2015optimizing, costa2016age, bedewy2016optimizing, kam2018age, yates2018age, inoue2019general, bedewy2019minimizing, bedewy2019age, najm2019content, yates2020age, kosta2017isit, kosta2020cost, moltafet2020age, najm2022optimal}). Reviews of AoI analysis in queueing networks can be found in~\cite{kosta2017age, yates2021age, sun2022age}. It has been observed that the last-come, first-served (LCFS) policy can achieve a smaller average age than several other queueing policies, such as first-come, first-served (FCFS). This aligns with the intuition that, in systems where the freshest packet best reflects the current system state, discarding outdated packets from the queue helps maintain real-time status updates and timely operation.

Numerous studies have developed age-optimal status updating policies for wireless networks under resource constraints, such as limited channel bandwidth and energy budgets. A recent review of such applications in 5G and beyond systems can be found in~\cite{pappas2023age}. The optimal policies for minimizing AoI with an energy-harvesting sensor were derived in, e.g.,~\cite{yates2015lazy, wu2017optimal, farazi2018average, gindullina2021age, farazi2018age, abd2022age, feng2021age, fang2022age, zheng2019closed, chen2019age, arafa2019age, hatami2021aoi}. Moreover, age-optimal scheduling of wireless networks has been investigated in, e.g.,~\cite{sun2017update, he2017optimal, kadota2018scheduling, kadota2019scheduling, joo2018wireless, hsu2019scheduling, lu2018age, sun2021age, zhou2019joint, talak2020improving, talak2018optimizing, tripathi2024whittle, ceran2019average}. Emulations and measurements of age-based policies were conducted in~\cite{kam2015experimental, kam2017modeling, tripathi2023wiswarm, ayan2021experimental, kutsevol2023experimental, shreedhar2024acp+}. It has been shown that age-optimal scheduling policies typically exhibit threshold-type structures; that is, communication is triggered whenever the age exceeds a fixed threshold value. Various works have considered decentralized (uncoordinated) status updates in multiaccess networks~\cite{yates2017status, jiang2018can, sun2019closed, leng2019age, jiang2019timely, maatouk2020csma, chen2020age, yavascan2021analysis, chen2022age, wang2024analytical, zhao2024age}. Motivated by the optimality of age-threshold policies, \cite{sun2019closed, yavascan2021analysis, zhao2024age} developed age-threshold random access protocols. Moreover, recent studies have developed pull-based policies for AoI optimization~\cite{chiariotti2022query, ildiz2023pull, stamatakis2024semantics, agheli2025}.

Recent studies have explored the role of AoI in NCSs. In the remote estimation and control of linear Gaussian systems, the estimation error covariance is a monotonic, nonlinear function of AoI (see, e.g.,~\cite{wu2013event, leong2017sensor, wu2020learning, ayan2019age, sun2019jcn, liu2022remote, jiping2025TACsubmission}). This reveals the connection between age and distortion in such systems. However, this may not always be the case, as AoI ignores the source evolution and application context. Not surprisingly, age-based policies are often suboptimal in monitoring general stochastic processes such as Wiener processes~\cite{sun2019sampling, pan2023sampling, tang2022sampling}, Ornstein-Uhlenbeck processes~\cite{ornee2021sampling, wang2022framework}, and Markov chains~\cite{stamatakis2019control, pappas2021goal, maatouk2020age, maatouk2022age, salimnejad2023state, salimnejad2024real, luo2025semantic, luo2025cost}. In general, the relationship between age and distortion is not well understood. Several recent works have studied the tradeoff between age and distortion~\cite{kam2018towards, inan2021optimal, bastopcu2021age, jayanth2023distortion}. 

\subsection{Information Semantics} 
The age is typically measured at the point of reception\footnote{The Age of Actuation (AoA)~\cite{nikkhah2023age, nikkhah2024age} addresses this limitation by measuring the timeliness of information at the point of actuation.}. However, fresh measurements may still have limited value for subsequent decision-making or actuation processes. Beyond accuracy and freshness, semantics-aware communications aim to \emph{assess the value (semantics) of information to generate and transmit only the relevant information at the right time}~\cite{howard1966information, juba2008universal, bisdikian2013quality, kountouris2021semantics, uysal2022Network, gunduz2023beyond, utkovski2023semantic}. A fundamental challenge is to extract semantic attributes and mathematically formalize value of information measures to guide the generation, transmission, reconstruction, and utilization of information for specific tasks or goals.

Several AoI variants have been introduced to balance freshness and correctness. For example, the Effective Age~\cite{kam2018towards} penalizes the cumulative error in the absence of updates. Similar in spirit, the Age of Incorrect Information (AoII) and the Cost of Memory Error~\cite{salimnejad2023state} depend on the duration for which the system remains in erroneous states. Early studies on these metrics focused on symmetric or binary Markov chains~\cite{kam2020age, maatouk2020age, maatouk2022age}. These results were extended to systems with random channel delays~\cite{chen2024minimizing}, retransmissions~\cite{bountrogiannis2024age}, multiple access~\cite{nayak2023decentralized, hong2023age, munari2024monitoring, AgheliCOMML24, chiariotti2025distributed}, continuous-time sources~\cite{cosandal2024modeling, cosandal2025multi}, query-based sampling~\cite{cosandal2024modeling, cosandal2024joint}, autoregressive Markov processes~\cite{joshi2021minimization}, among others.

For applications with slowly evolving information sources, freshness can be measured in terms of updates rather than timestamps to avoid unnecessary growth in age. The authors of~\cite{zhong2018two} introduced the Age of Synchronization (AoS) for caching systems, where the age remains zero when there is no update at the source. Binary freshness is another variant, which concerns whether the updates requested from the cache are up-to-date~\cite{bastopcu2020maximizing, bastopcu2020information, kaswan2021freshness, akar2024query}. Similarly, the Version Age of Information (VAoI)~\cite{yates2021Vage} measures information freshness in networks and freezes the growth in age when there is no new update (i.e., a version) at the source. Analysis of VAoI in gossiping and multi-hop networks was reported in~\cite{buyukates2022version, abd2023distribution, kaswan2024timestomping, mitra2024age, kaswan2025misinformation, kaswan2025age, delfani2025timestamps}. VAoI-optimal status updating policies for energy-harvesting sensors and satellite systems were derived in~\cite{delfani2024qvaoi, delfani2024version, Delfani2025LEO}. Version age has also been applied to the remote estimation of Markov sources~\cite{champati2022detecting, salimnejad2025age}.

The above metrics, however, treat all information content equally and do not explicitly consider the \emph{contextual relevance} and goal-oriented usefulness. Information that is both fresh and accurate may still be of limited value if it does not contribute to the ultimate goal. Early efforts to address this limitation include the content-aware AoI~\cite{stamatakis2019control, delfani2025state}, which models the information source as a Markov chain with both normal and alarm (urgent) states. This metric tracks the staleness of each state's information separately. It reveals that the quality of information depends on its content and evolves at different rates. Similar considerations can be found in uncertainty of information~\cite{chen2022uncertainty, chen2023index, cocco2023remote}, and age of channel state information~\cite{costa2015ISIT, costa2015ICC, lipski2024age, OnurPIMRC2025, chakraborty2025send}.

Context-aware status updates in NCSs have gained significant interest in recent years. The Cost of Actuation Error (CAE)~\cite{pappas2021goal} captures the fact that the cost of estimation error depends not only on the physical discrepancy but also on the contextual relevance and potential control risks to the system’s performance. Follow-up studies in~\cite{salimnejad2024real, salimnejad2023state} derived analytical results for monitoring Markov chains. Optimal policies for minimizing CAE were developed in~\cite{fountoulakis2023goal, zakeri2024semantic, luo2024goal, ornee2023context, luo2025semantic}. In particular, \cite{luo2024goal, luo2025semantic} developed optimal and learning-based communication policies in resource-constrained multi-source systems, a setting relevant for multimodal scenarios. Another related metric is the Urgency of Information (UoI)~\cite{zheng2020urgency}, which penalizes information staleness using context-aware weights. Moreover, \cite{holm2023goal} examined a scenario in which the end user requests different statistical properties of a dynamical system depending on the application context. These metrics rely solely on attributes extracted from the current system state, whereas the history-dependent value and implications are ignored. 

History matters---a philosophy common in science, culture, and human society. Information value depends on the history of all system realizations and decisions. An important semantic attribute motivated by many NCSs is the \emph{lasting impact} in consecutive estimation errors. That is, the longer an error persists, the more severe its consequences can become~\cite{salimnejad2023state, luo2025cost}. The AoII~\cite{maatouk2020age} captures this notion, though it treats all errors equally and is therefore context-agnostic. This egalitarianism leads to inadequate transmissions in urgent states but excessive transmissions in normal states. Motivated by this, \cite{luo2024exploiting, luo2024minimizing} introduced the Age of Missed Alarm (AoMA) and the Age of False Alarm (AoFA) to quantify, respectively, the persistence costs in the more urgent missed alarm error and the less important false alarm error. Similar metrics were also studied in the context of health monitoring in~\cite{bastopcu2022using}. Moreover, the severity of an estimation error depends on both its holding time and context. The significance-aware Age of Consecutive Error (AoCE)~\cite{luo2025cost} thus measures the \emph{urgency of lasting impact} using a collection of context-aware nonlinear functions. It has been shown that the optimal policy has a simple switching structure whose transmission thresholds depend on the instantaneous estimation error~\cite{luo2025cost}. The follow-up study in~\cite{luo2025role} developed optimal policies to minimize AoCE with a maximum a posteriori (MAP) estimator whose predictability is characterized by AoI. This work reveals the connection between age and the semantics of information in such systems.

\subsection{Analytical Tools}
This section reviews key analytical tools used to control information generation and transmission processes. With proper value of information measures at hand, these tools determine when to generate updates and how to transmit them efficiently, i.e., the \emph{value of timing} problem. \emph{The optimal timing depends not only on the information value extracted from past observations but also on predictions of future realizations derived from the system dynamics.} This dual reliance on past and predicted information introduces significant computational and analytical challenges compared with the classical design of distortion- or delay-optimal systems. Depending on how much predicted information is incorporated into planning, these analytical tools can be classified as \emph{myopic}, \emph{finite-horizon}, or \emph{full-horizon} approaches.

Myopic methods rely solely on the current system state, without incorporating any predictions or foresight about future realizations. Randomized, reactive, and periodic policies fall into this category. Although these policies are often suboptimal, they are easy to implement and serve as the basis for system modeling and evaluation. For example, randomized stationary policies offer a simple yet tractable approach to analyzing long-term system performance and can yield closed-form expressions for metrics such as average age; see, e.g.,~\cite{bastopcu2020information, salimnejad2024real, salimnejad2023state, akar2024query}.

Finite-horizon approaches consider possible trajectories over multiple time steps. Two dominant methods in the communication and control literature are the Lyapunov optimization (which typically incorporates one-step lookahead)~\cite{neely2010stochastic} and Model Predictive Control (MPC)~\cite{MPC}. A salient feature of these methods is their computational efficiency: they require no offline computation and can adjust their behavior online by adapting to the current system state and short-term predictions~\cite{kadota2018schedluing, kadota2019scheduling, fountoulakis2023scheduling, ramakanth2025monitoring}. Although the Lyapunov approach does not provide structural insight into the optimal policy or its transmission schedule, it offers an effective means of handling hard constraints in the optimization of large-scale systems~\cite{luo2025semantic, cui2012survey, zakeri2023minimizing}.

Full-horizon approaches, primarily based on Markov decision processes (MDPs)~\cite{puterman1994markov, altman2021constrained, krishnamurthy2016POMDP}, play a central role in goal-oriented semantic communications. MDPs provide a systematic framework for characterizing the structure and behavior of optimal policies, enabling the determination of optimal timing and the best achievable performance (see, e.g.,~\cite{leong2017sensor, sun2019sampling, wang2020minimizing, tang2020minimizing, luo2025cost, cosandal2025multi, agheli2025, talli2025pragmatic}). However, this level of optimality comes at the cost of high computational complexity, particularly for high-dimensional sources or large-scale systems~\cite{luo2025semantic}. To mitigate this complexity, low-complexity approximations such as Lyapunov-based, token-based, and Whittle's index-based policies have been developed (see, e.g.,~\cite{ stamatakis2023optimizing, delfani2025optimizing, whittle1988restless, liu2010indexability, hsu2019scheduling, sun2020closed, chen2023index, tripathi2024whittle}). When the system model is not known a priori, model-free approaches such as reinforcement learning (RL) can be applied to learn near-optimal policies from interaction with the environment~\cite{sutton1998reinforcement, abd2020reinforcement, ceran2019average, wu2020learning, chen2024structure, luo2025semantic}.

\section{Goal-Oriented Semantics-Aware Communication Paradigm}
This section outlines a high-level architecture for goal-oriented, semantics-aware communication. We begin by characterizing information sources that capture a broad range of applications. We then introduce a point-to-point setup that illustrates the full chain of information generation, transmission, reconstruction, and goal-driven utilization, followed by a generalized framework that extends these ideas to networked intelligent systems. We conclude with key application domains where semantics-aware communication is expected to play a key role in 6G and beyond.

\subsection{Information source}
Here, we discuss the types of information sources that generate data to be transmitted in the network. We primarily focus on independent and identically distributed (i.i.d.) source models and Markov chains. We next demonstrate that these models are relevant and capture a wide range of applications.

Markov chains are powerful tools for modeling temporal \emph{correlation} and state \emph{uncertainty} in dynamical systems. Owing to their simplicity and rich structure, they are ubiquitous in modern intelligent systems for sequential reasoning, prediction, and planning. In artificial intelligence (AI), Markov chains underpin reinforcement learning through MDPs~\cite{puterman1994markov}, enabling agents to make informed decisions under uncertainty~\cite{sutton1998reinforcement}. Inspired by the Markovianity of natural languages (e.g., Shannon~\cite{shannon, shannon1951prediction}), recent studies have used Markov chains as a theoretical lens to systematically analyze the reasoning capabilities of neural networks~\cite{buesing2011neural, makkuva2024attention}. In control systems, they are widely used to model disturbances and anomalies, allowing for robust control and fault-tolerant system design~\cite{costa2005mjls}. In wireless communications, they capture system uncertainties such as channel fading, user mobility, and traffic patterns, enabling more adaptive and reliable network configurations~\cite{hong1995finite, ephraim2002hidden}. 

Across these applications, the states in a Markov chain can represent either quantized levels of a physical process (e.g., channel condition or system disturbance) or abstract system status (e.g., operating mode or anomaly). For example, an industrial plant may shift into an abnormal mode due to environmental disturbances or anomalies. This information is often hidden but can be inferred from noisy measurements~\cite{costa2005mjls}. Consequently, Markov chains can encode the contextual relevance of information as well as the potential control costs or risks associated with delayed access to critical information~\cite{luo2025cost}. High-dimensional Markov chains thus provide a systematic representation of multimodal systems with heterogeneous information flows of varying significance~\cite{luo2025semantic}.

I.i.d. models are special cases of Markov chains in which temporal correlation is absent. The vast majority of existing studies on AoI and its variants assume i.i.d. source models (see, e.g.,~\cite{kosta2017age, yates2021age, buyukates2022version}). In control theory, the derivation of Kalman filters and optimal control strategies relies on the common assumption of i.i.d. noise and disturbances~\cite{brian1979optimal}.


\subsection{From Point-to-Point Communication to a Semantics-Aware Information Chain}
In this paper we move beyond the conventional point-to-point communication paradigm. We consider the entire information chain, from the moment information is generated through sensing or extracted from measurements, to its transmission over a communication network, and ultimately to its reconstruction and utilization at a remote location as illustrated in Fig.~\ref{fig_HighLevelSystem}. This approach enables additional degrees of freedom for assessing the importance and utility of information. By jointly considering all the stages of the information lifecycle, the system can more effectively determine when information is redundant, outdated, or irrelevant, thereby reducing unnecessary transmissions and enabling more efficient use of communication and computational resources. Such an approach redefines communication not as an isolated process, but as a coordinated component of a broader goal-oriented system.

This expanded perspective also allows the exploitation of information properties at different stages of the chain. As proposed in~\cite{kountouris2021semantics, pappas2021goal}, information attributes can be decomposed into \emph{innate} and \emph{contextual}, each modelled by relevant metrics that capture their semantic significance. Throughout this paper, we adopt a Markovian abstraction of the information source, which provides a general yet analytically tractable model for capturing temporal correlations, state transitions, and context-dependent relevance as described earlier. The objective is to accurately reconstruct and utilize information from this source at a remote destination for actuation, where actuation may be physical or virtual, including any form of information-driven decision-making or control. This viewpoint emphasizes that semantics-aware communication is not only about transmitting data, but about enabling goal-oriented and effective actions based on that data. We will investigate those in more detail in Sections \ref{sec:semantics} and \ref{sec:solution}.

\subsection{General Setup}
A more general and realistic setting moves beyond the simple point-to-point case and considers environments where many physical or digital processes operate simultaneously and influence one another. These processes act as distributed information sources, while monitoring devices or agents observe them, extract relevant features, and prepare the information for transmission. Such processes may be multidimensional, correlated, or multimodal, and in many cases only partially observable. Their evolution may also depend on closed-loop interactions, where remote decisions or actuations can shape how the system behaves over space and time. A representative example is that of two agents collaborating to complete a shared task under strict energy and time constraints, where each agent’s actions and transmitted information affect the other’s performance. Furthermore, these semantics-aware goal-oriented information exchanges often coexist with conventional traffic, such as video streams or background data, reinforcing the need for communication strategies that allocate resources based on the actual value of the information rather than treating all data equally.

\subsection{Potential Applications}
A wide range of emerging systems stand to benefit substantially from semantics-aware and goal-oriented communication as depicted in Fig. \ref{fig_HighLevelSystem}. In robotics and autonomous systems, timely and context-relevant information exchange is critical for safe navigation, cooperative manipulation, and collective decision-making. Robots frequently operate under tight energy and bandwidth constraints, yet must react to dynamic environments. By prioritizing only important information, semantics-aware goal-oriented communication enables more efficient coordination and reduces unnecessary updates within multi-robot teams. Similarly, in industrial automation and smart manufacturing, networked sensors and controllers must monitor fast-evolving processes in which abnormal events pose disproportionate operational risks. Integrating semantic metrics enables these systems to filter redundant status updates, focus communication resources on deviations that affect safety or production quality, and achieve reliable, real-time control despite limited communication budgets.

Digital twins and large-scale cyber-physical replicas also provide strong motivation for goal-oriented information exchange. Maintaining an accurate virtual representation requires efficient synchronization with the physical system; thus, transmitting all measurements at scale is not practical or even feasible. Semantics-aware goal-oriented communication enables selective synchronization based on the value of updates to prediction accuracy, anomaly detection, or operational planning. Satellite networks and spaceborne sensing platforms constitute another domain where semantic principles are essential. Constrained by limited power, intermittent connectivity, and long propagation delays, these systems must decide when and what to transmit. By leveraging semantic significance, for instance, prioritizing relevant observations, critical changes, or mission-essential state information, it can significantly improve data utility while reducing communication load. Across these diverse applications, semantics-aware and goal-oriented communication provides a principled approach to achieving sustainable, efficient, and mission-driven information exchange. A more detailed treatment can be found in Section \ref{sec:applications}.

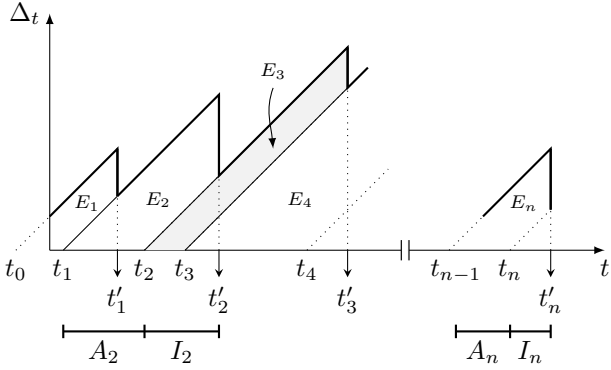
\begin{figure}[t!]
    \centering
    \begin{tikzpicture}[scale=0.9]
\draw[->] (0,0) -- (8.2,0) node[anchor=north] {$t$};
\draw[->] (0,0) -- (0,3.5) node[anchor=east] {$\Delta_t$};

 \draw[fill=gray!10] (1.4,0) -- (4.4,3) -- (4.4,2.4)-- (2,0);

\draw	
(-0.5,0) node[anchor=north] {$t_0$}
(0.2,0) node[anchor=north] {$t_1$}
(1.4,0) node[anchor=north] {$t_2$}
(2,0) node[anchor=north] {$t_3$}
(3.8,0) node[anchor=north] {$t_4$}
(6,0) node[anchor=north] {$t_{n-1}$}
(6.8,0) node[anchor=north] {$t_n$};
		    
\draw[->,>=stealth] (1,0) -- (1,-0.4) node[anchor=south,below] {$t'_1$};
\draw[->,>=stealth] (2.5,0) -- (2.5,-0.4) node[anchor=south,below] {$t'_2$};
\draw[->,>=stealth] (4.4,0) -- (4.4,-0.4) node[anchor=south,below] {$t'_3$};
\draw[->,>=stealth] (7.4,0) -- (7.4,-0.4) node[anchor=south,below] {$t'_n$};
   		 
\draw 
(0.55,0.7) node{{\scriptsize $E_1$}}
(1.6,0.75) node{{\scriptsize $E_2$}}
(3.7,0.75) node{{\scriptsize $E_4$}}
(7.0,0.7) node{{\scriptsize $E_n$}};
           
\draw[<-] (3.3,1.5) to [out=95,in=250] (3.3,2.4) node [above] {{\scriptsize $E_3$}};           
\draw [thick](0.2,-1.2) -- (1.4,-1.2) node[pos=.5,sloped,below] {$A_2$} ;
\draw[thick] (0.2,-1.3) -- (0.2,-1.1); 
\draw[thick] (1.4,-1.2) -- (2.5,-1.2) node[pos=.5,sloped,below] {$I_2$} ;
\draw[thick] (1.4,-1.3) -- (1.4,-1.1) 
             (2.5,-1.3) -- (2.5,-1.1);
                    
\draw[thick] (6,-1.2) -- (6.8,-1.2) node[pos=.5,sloped,below] {$A_n$} ;
\draw[thick] (6,-1.3) -- (6,-1.1); 
\draw[thick] (6.8,-1.2) -- (7.4,-1.2) node[pos=.5,sloped,below] {$I_n$} ;
\draw[thick] (6.8,-1.3) -- (6.8,-1.1) 
             (7.4,-1.3) -- (7.4,-1.1);
 
\draw[thick] (0,0.5) -- (1,1.5) -- (1,0.8);
\draw (0.2,0) -- (1,0.8);
\draw[thick] (1,0.8) -- (2.5,2.3) -- (2.5,1.1) -- (4.4,3) -- (4.4,2.4) -- (4.7,2.7);
\draw[thick] (6.4,0.5) -- (7.4,1.5) -- (7.4,0.6);

\draw[thick] (1,0.8) -- (1,1.5);
\draw[thick] (2.5,1.1) -- (2.5,2.3);
\draw[thick] (4.4,2.4) -- (4.4,3);
\draw[thick] (7.4,0.6) -- (7.4,1.5);

\draw[dotted] (1,0) -- (1,0.8);
\draw[dotted] (2.5,0) -- (2.5,1.1);
\draw[dotted] (4.4,0) -- (4.4,2.4);
\draw[dotted] (7.4,0) -- (7.4,0.6);
\draw[dotted] (-0.5,0) -- (0,0.5);
\draw[dotted] (1.4,0) -- (2.5,1.1);
\draw[dotted] (2,0) -- (4.7,2.7);
\draw[dotted] (3.8,0) -- (5,1.2);
\draw[dotted] (5.9,0) -- (6.4,0.5);
\draw[dotted] (6.8,0) -- (7.4,0.6);

\draw[thick] (5.2,-0.15) -- (5.2,0.15) 
            (5.3,-0.15) -- (5.3,0.15);
\draw[white, fill=white!50] (5.21,-0.2) -- (5.21,0.2) -- (5.29,0.2) -- (5.29,-0.2) ;                   
\end{tikzpicture}
    \caption{AoI and its sawtooth pattern~\cite{kaul2012real}. Status updates are generated at time $\{t_n\}_{n\geq 1}$ and will be received at the receiver at time $\{t_n^\prime\}_{n\geq 1}$.}
    \label{fig:linear age}
\end{figure}

\section{Freshness of Information}
\subsection{Information Freshness and AoI}
AoI quantifies the freshness of the knowledge that a remote receiver has about an information source. At any time $t$, if the latest received update at the destination (e.g., a remote monitor or controller) was generated at time $G_t \leq t$, then the AoI at the destination is defined as
\begin{align}
    \Delta_t = t - G_t.
\end{align}

Consider a status update system where a sequence of packets is generated at rate $\lambda$ at times $t_1, t_2, t_3, \ldots$, and is successfully delivered to the receiver at times $t_1^\prime, t_2^\prime, t_3^\prime, \ldots$. Fig.~\ref{fig:linear age} illustrates the evolution of AoI as a function of time $t$. As observed, the AoI increases linearly until the reception of an update, at which point the age drops to a smaller value $I_i = t^\prime_i - t_i$, where $I_i$ is the total time the $i$th packet spent in queueing and service. Let $A_i = t_i - t_{i-1}$ denote the inter-arrival times. The average AoI, which corresponds to the average area of the $E_i$ trapezoids (i.e., the shaded areas) in Fig.~\ref{fig:linear age}, can be calculated as~\cite{kaul2012real}
\begin{align}
    \bar{\Delta} = \lambda(\mathbb{E}[A I]+\mathbb{E}[A^{2}]/2),
\end{align}
where $\mathbb{E}[\cdot]$ is the expectation operator, and $A$ and $I$ are the random variables corresponding to the inter-arrival time and the system time of an update packet, respectively. For FCFS M/M/1 and M/D/1 queues with arrival rate $\lambda$ and service rate $\mu$, the average AoI is obtained as~\cite{yates2021age}
\begin{align}
    \bar{\Delta}_{\text{M/M/1}} &= \frac{1}{\mu} \Big(1 + \frac{1}{\rho} + \frac{\rho^2}{1 - \rho}\Big),\\
    \bar{\Delta}_{\text{M/D/1}} &= \frac{1}{\mu} \Big(\frac{1}{2(1-\rho)} + \frac{1}{\rho} + \frac{(1-\rho)\mathrm{exp}(\rho)}{\rho}\Big),
\end{align}
where $\rho = \lambda/\mu$ is the offered load. Moreover, the throughput is $\lambda$, and the average delay is $\mathbb{E}[I] = 1 / (\mu - \lambda)$.

\begin{figure}[t!]
    \centering
    \begin{tikzpicture}
\begin{axis}[
    width=9cm, height=4.5cm,
    xlabel={\text{Offered load} $\rho$},
    ylabel={\text{Age} $\bar{\Delta}$},
    xmin=0.1, xmax=0.9,
    ymin=2, ymax=8,
    ytick={2,4,6,8},
    samples=200,
    domain=0.1:0.9,
    thick,
    legend style={
        at={(0.5,0.95)}, 
        anchor=north,
        draw=black, 
        fill=white, 
    },
]

\addplot[
    blue, thick,
    mark=o,
    mark options={draw=black, fill=none},
    mark repeat=10, 
] {1 + 1/x + (x^2)/(1-x)};
\addlegendentry{$\text{M/M/1}$}

\addplot[
    red, thick,
    mark=square, mark options={draw=black, fill=none},
    mark repeat=10
] {1/(2*(1 - x)) + 1/2 + (1 - x)*exp(x) / x};
\addlegendentry{$\text{M/D/1}$}

\end{axis}
\end{tikzpicture}
    \caption{Average age versus offered load $\rho$ for FCFS M/M/1 and M/D/1 queueing systems with service service $\mu = 1$~\cite{kaul2012real}.}
    \label{fig:agedelay}
\end{figure}
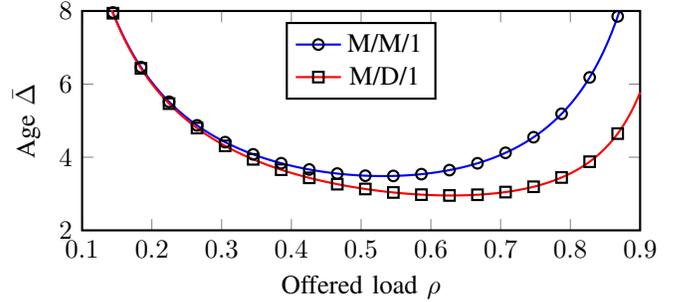

Fig.~\ref{fig:agedelay} illustrates the average AoI for FCFS M/M/1 and M/D/1 queues with a service rate of $\mu = 1$. In such settings, the maximum throughput is achieved as $\rho \to 1$; that is, the sensor generates as much information as possible to keep the server busy. In contrast, the minimum packet delay occurs as $\rho \to 0$, when very few packets are generated so they spend less time in the queue. However, optimizing solely for delay or throughput can significantly degrade the system's AoI performance. This highlights the need for careful calibration of the information generation rate to optimize freshness. A detailed treatment of these topics can be found in several surveys and books~\cite{kosta2017age, sun2022age, yates2021age, pappas2023age}.

In the next subsection we deal with the following question: Does information staleness necessarily grow linearly with time? 

\begin{figure}[t!]
    \centering
    \begin{tikzpicture}[scale=0.9]
\draw[->] (0,0) -- (8.2,0) node[anchor=north] {$t$};
\draw[->] (0,0) -- (0,3.5) node[anchor=east] {$f(\Delta_t)$};

\draw[fill=gray!10] (1.4,0) to[bend right=20] (4.4,3) -- (4.4,2.1) to[bend left=17] (2,0);
\draw 
(-0.5,0) node[anchor=north] {$t_0$}
(0.2,0) node[anchor=north] {$t_1$}
(1.4,0) node[anchor=north] {$t_2$}
(2,0) node[anchor=north] {$t_3$}
(3.8,0) node[anchor=north] {$t_4$}
(6,0) node[anchor=north] {$t_{n-1}$}
(6.8,0) node[anchor=north] {$t_n$};

\draw[->,>=stealth] (1,0) -- (1,-0.4) node[anchor=south,below] {$t'_1$};
\draw[->,>=stealth] (2.5,0) -- (2.5,-0.4) node[anchor=south,below] {$t'_2$};
\draw[->,>=stealth] (4.4,0) -- (4.4,-0.4) node[anchor=south,below] {$t'_3$};
\draw[->,>=stealth] (7.4,0) -- (7.4,-0.4) node[anchor=south,below] {$t'_n$};
		    	    		 
\draw (2,0.8) node{{\scriptsize $E_2$}};
\draw (4.0,0.85) node{{\scriptsize $E_4$}}
      (7,0.4) node{{\scriptsize $E_n$}};   
\draw[<-] (0.44,0.33) to [out=95,in=250] (0.44,1.1) node [above] {{\scriptsize $E_1$}};                        
\draw[<-] (3.35,1.0) to [out=95,in=250] (3.35,1.9) node [above] {{\scriptsize $E_3$}};                
 
\draw[thick] (1,1.5) -- (1,0.45);
\draw[thick] (2.5,2.3) -- (2.5,0.6);
\draw[thick] (4.4,3) -- (4.4,2.1);
\draw[thick] (7.4,1.5) -- (7.4,0.3);

\draw[dotted] (-0.5,0) to[bend right=20] (1,1.5); 
\draw[thick] (0,0.27) to[bend right=15] (1,1.5); 

\draw (0.2,0) to[bend right=10] (1,0.45); 
\draw[thick] (1,0.45) to[bend right=15] (2.5,2.3); 

\draw[dotted] (1.4,0) to[bend right=20] (4.4,3); 
\draw[thick] (2.5,0.6) to[bend right=15] (4.4,3.05); 

\draw[dotted] (2,0) to[bend right=20] (4.7,2.7); 
\draw[thick] (4.4,2.1) to[bend right=3] (4.7,2.7); 

\draw[dotted] (5.9,0) to[bend right=20] (7.4,1.5); 
\draw[thick] (6.4,0.27) to[bend right=17] (7.4,1.5); 

\draw[dotted] (1,0) -- (1,0.5);
\draw[dotted] (2.5,0) -- (2.5,2);
\draw[dotted] (4.4,0) -- (4.4,2.4);
\draw[dotted] (7.4,0) -- (7.4,0.6);

\draw[dotted] (3.8,0) to[bend right=20] (5,1.2);
\draw[dotted] (6.8,0) to[bend right=20] (7.4,0.3);

\draw[thick] (5.2,-0.15) -- (5.2,0.15) 
            (5.3,-0.15) -- (5.3,0.15);
\draw[white, fill=white!50] (5.21,-0.2) -- (5.21,0.2) -- (5.29,0.2) -- (5.29,-0.2) ;                               
\end{tikzpicture}
    \caption{Sample path of a non-linear age function. New updates are generated at time $\{t_n\}_{n\geq 1}$ and will be received at the receiver at time $\{t_n^\prime\}_{n\geq 1}$.}
    \label{fig:exponential_value}
\end{figure}
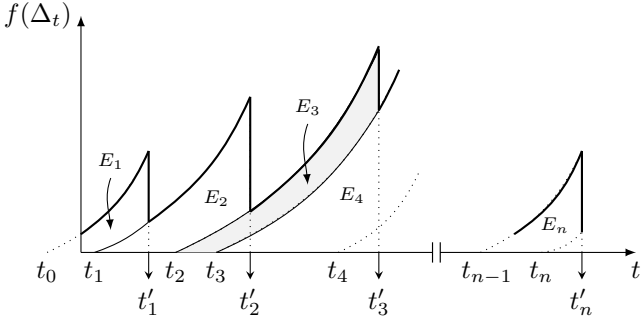

\subsection{Nonlinear Aging}\label{sec:nonlinear-aging}
While linear age metrics are often assumed, real-world applications often exhibit nonlinear relationships between staleness and utility~\cite{kosta2017isit, sun2019jcn, shisher2024timely, jiping2025TACsubmission}. As depicted in Fig.~\ref{fig:exponential_value}, the cost of having outdated status updates can be modeled as a nonlinear function $f(\Delta_t)$, where $f: \mathbb{N}\to[0,\infty)$ is a monotonically non-decreasing function. 

The following canonical example in NCSs illustrates the notion of nonlinear AoI. Consider an NCS in which a sensor observes a linear Gaussian system
\begin{align}
    X_{t+1} = A X_t + W_t, \label{eq:linear-gaussian-source}
\end{align}
where $X_t \in \mathbb{R}^{n}$ is the system state at time slot $t$, $A \in \mathbb{R}^{n \times n}$ is the state transition matrix, and $W_t \in \mathbb{R}^{n}$ is zero-mean Gaussian process noise. The sensor measurements are given by $Y_t = C X_t + V_t$, where $C \in \mathbb{R}^{m \times n}$ is the measurement matrix, and $V_t \in \mathbb{R}^m$ is zero-mean Gaussian measurement noise. The sensor runs a steady-state Kalman filter to pre-estimate the source state from the history of all measurements $Y_{1:t}$. At each time $t$, the sensor produces a local estimate $\hat{X}_t^\text{local}$ with a \emph{constant} error covariance $\bar{P}$. In this setting, all generated information is equally accurate, and the freshest estimate best reflects the system state. Hence, transmitting outdated packets is wasteful if a newer measurement is available.

In practical sensor networks, the sensor can transmit only intermittently over an unreliable wireless channel. Therefore, the receiver must reconstruct the source state using intermittent and potentially outdated information. Suppose the receiver's latest received measurement was generated at time $G_t<t$, and hence the AoI is $\Delta_t = t-G_t$. At time $G_t$, the receiver's estimate is equal to the sensor's local estimate at that time, i.e., $\hat{X}_{G_t} = \hat{X}^\text{local}_{G_t}$. Since no updates have been received within the interval $[G_t, t]$, at time $t$ the receiver has to construct an estimate of the current source state using the outdated information. It has been shown that the estimation error covariance at time $t$ is a monotonic, non-decreasing, and nonlinear function of AoI~\cite{wu2020learning, jiping2025TACsubmission}, i.e., 
\begin{align}
    \mathbb{E}[(X_t - \hat{X}_t)(X_t - \hat{X}_t)^\top] = f(\Delta_t), 
\end{align}
where $f$ is determined by the system statistics and noise distribution. This nonlinear relationship can be generalized to higher-order linear sources of the form \cite{chowdhury2024on} 
\begin{align}
    X_t = a_1 X_{t-1} + a_2 X_{t-2} + \ldots + a_{L} X_{t-L} + W_t,
\end{align}
provided that the sensor transmits a sequence of measurements no shorter than the source order $L$.

\subsection{Shortcomings of Freshness Metrics}
\label{Sec_Shortcomings}
Age is an innate, application-independent attribute of information. However, it overlooks several critical aspects that are essential for effective communication.

Age is typically measured at the point of reception. In many NCSs, however, the receiver must feedback control commands based on the received status update. In such cases, the usefulness of information depends not only on its freshness but also on its relevance to the control task and the timing of its utilization. Updates that arrive when the system cannot act on them, or when the state has not changed significantly, may provide little value for their operational cost, yet they affect the AoI metric. Early efforts to address this limitation include the AoA and Age of Actuated Information (AoAI)~\cite{nikkhah2023age, nikkhah2024age}, which extends the information chain to account for the utilization and actuation of received updates. 

Moreover, age alone may not suffice to characterize the value or effectiveness of information, particularly in scenarios where content significance, contextual relevance, or task-specific requirements play a key role. In real-world applications, the value of information is rarely determined by a single attribute; instead, it requires a holistic measure that combines multiple factors, including timeliness, relevance, correctness, and utility. This motivates the development of systematic, semantics-aware metrics to guide the generation, transmission, reconstruction, and utilization of information.

\section{Semantics of Information beyond AoI}\label{sec:semantics}
To motivate the notion of information semantics, we first discuss, from both control- and information-theoretic perspectives, that information accuracy and freshness alone may not suffice for 6G and beyond systems.

In classical information theory, the primary objective is to quantify the \emph{amount} of information based only on the
probabilistic structure of the source~\cite{shannon}, while the \emph{significance} and \emph{timing} aspects of information are generally ignored~\cite{howard1966information, anantharam1996bits, popovski2022perspective}. A fundamental problem is the rate-distortion tradeoff, which addresses the question of ``how accurately the symbols of communication can be transmitted". However, this paradigm faces challenges in emerging data-intensive cyber-physical systems, where the effectiveness of communication is evaluated by its impact on system performance rather than solely on signal fidelity.

In classical control theory, a fundamental problem is the remote estimation of stochastic processes. Estimation quality is traditionally measured using distortion measures such as Hamming distortion or mean squared error~\cite{schenato2007foundations, hespanha2007survey}, where a measurement is considered valuable if its reception improves the accuracy of the receiver's estimate. The underlying assumption is that all source states convey equally important information, and the cost of an estimation error depends on the discrepancy between the source and the reconstructed signal. However, this assumption warrants careful re-examination in many applications.

As a motivating application, consider a connected autonomous vehicle navigating toward its target area following instructions from a remote control center, e.g., a cloud server or an edge computing unit~\cite{jiping2023TITS, chen2016cooperative, hult2016coordination}. The vehicle's onboard sensor determines when to transmit status updates (e.g., location, speed, acceleration) to the center to ensure safe maneuvering and efficient traffic management. On the one hand, due to the scarcity of communication resources and the immensity of sensory data, it is not possible to send every piece of sensory data. On the other hand, the raw measurements often contain correlated or redundant components that are of limited value for decision-making at the center. Naturally, we ask:
\begin{enumerate}
    \item Are all measurements equally important? Does more accurate or fresher mean more valuable?
    \item How should the sensor determine which measurements are valuable (i.e., the \textit{value of information})?
    \item When should the sensor report valuable measurements to the center (i.e., the \textit{value of timing})?
\end{enumerate}

Prior to providing the formal definitions and mathematical treatment to address the above questions, we first introduce and motivate key semantic attributes in an intuitive, nontechnical way. The following semantic attributes, drawn from the autonomous driving example, are of interest:
\begin{itemize}
    \item \textbf{Context-awareness:} Minor errors, such as slight lane deviations or misjudging the speed of a non-conflicting vehicle, may lead to suboptimal maneuvering and reduced driving comfort. In contrast, critical errors, such as failing to detect a nearby traffic participant, must be avoided entirely, as they can result in hazardous operations or even fatal accidents.
    \item \textbf{Cost of Consecutive Error (Lasting Impact):} If an error persists, its consequences may worsen as the system continues to operate under false assumptions. For example, if a vehicle consistently underestimates the distance to a lead car, the accumulated incorrect spacing over several seconds may force sudden braking or unsafe maneuvers, significantly increasing collision risk.
    \item \textbf{Urgency of Lasting Impact:} The severity of an error depends on both its duration and context. For instance, misjudging the speed of a non-conflicting vehicle may slightly reduce traffic throughput, whereas failing to detect a nearby traffic participant for even a few seconds could lead to a crash.
\end{itemize}

Similar considerations arise in many applications, including health monitoring, power grid management, financial trading, and anomaly detection in manufacturing systems. Thus motivated, goal-oriented semantic communications assess the value of information and aim to generate and transmit only what is relevant and at the right time. From the perspective of communication, control, and information theories, the need for semantic awareness has become increasingly evident and garnered significant attention in recent years. 

\subsection{Problem of Interest}\label{sec:problem-of-interest}
\begin{figure}[t!]
    \centering
    \includegraphics[width=\linewidth]{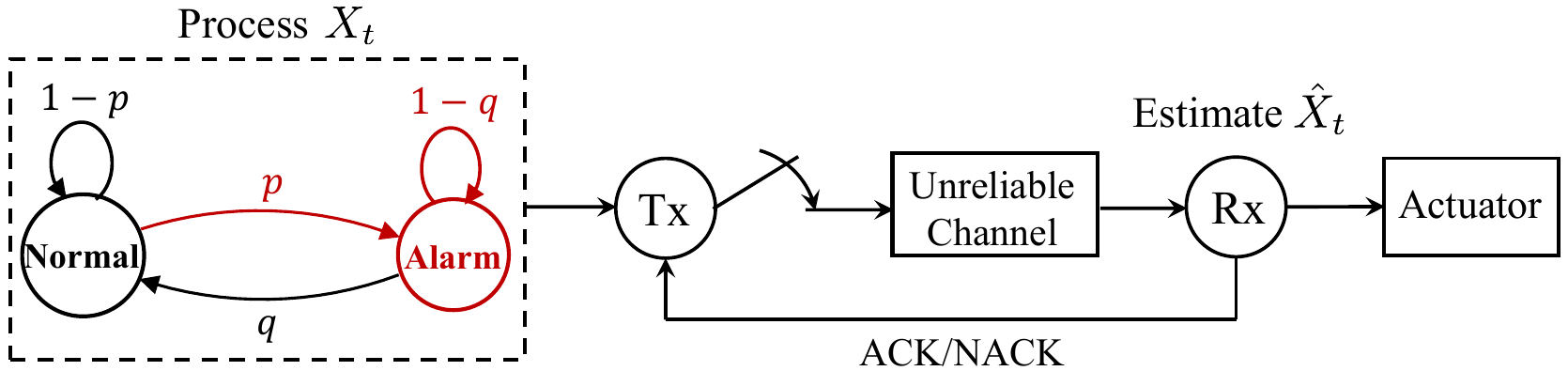}
    \caption{Remote estimation of a Markovian dynamical system for the purpose of actuation.}
    \label{fig:system-model-problem}
\end{figure}

Throughout this section, we will focus on the real-time tracking of discrete-state Markov sources, as depicted in Fig.~\ref{fig:system-model-problem}. This setup serves as the most indicative scenario for demonstrating semantic attributes beyond age and distortion, as well as operational schemes that jointly consider sampling, transmission, and information utilization.

\subsubsection{Source} The information source is modeled as a finite-state, time-homogeneous Markov chain $\{X_t\}_{t\geq 0}$, where $X_t\in\mathcal{X}$ is the state variable taking values in the finite set $\mathcal{X}$. For clarity, and unless stated otherwise, we illustrate the metrics using a binary Markov chain ($|\mathcal{X}| = 2$). We note that the metrics, and hence the analysis and results, can be extended to general finite-state source models. At each time $t$, the source resides in either state $0$ (\textit{normal}, low-priority) or state $1$ (\textit{alarm}, high-priority). The states here can represent either quantization levels of a physical process or the abstract status (e.g., operation modes, component failures, abrupt changes in system dynamics) of a system. We distinguish between two types of estimation errors:
\begin{itemize}
    \item \emph{Missed alarms} occur when the receiver declares the normal state while the source is actually in the alarm state, i.e., $X_t = 1, \hat{X}_t = 0$. Timely detection of abnormalities is crucial for decision-making and system maintenance.
    \item \emph{False alarms} occur when the receiver erroneously declares the alarm state while the source is in the normal state, i.e., $X_t = 0, \hat{X}_t = 1$. Though less critical, false alarms can lead to unnecessary expenditure on checking the system, thus wasting resources.
\end{itemize}

The state transition probability matrix $Q$ is given by
\begin{align}
    Q = \begin{bmatrix}
        1-p & p\\
        q & 1-q
    \end{bmatrix},
\end{align}
where $Q_{i,j} = \Pr[X_{t+1} = j|X_t = i]$ denotes the probability of transitioning from state $i$ to state $j$ between two consecutive time slots. To avoid pathological cases, we assume $Q$ is irreducible, i.e., $0<p,q<1$. A chain is called symmetric if $Q = Q^\top$; otherwise, it is called asymmetric. Moreover, the chain is called positively correlated if it satisfies $p + q < 1$, meaning that it is more likely to remain in the current state than to switch. The special case $p + q = 1$ corresponds to a source with i.i.d. state evolution.

\subsubsection{Sensor} The sensor observes the source and decides, at every decision epoch (i.e., the beginning of time slot $t$), whether to transmit a new measurement to the remote receiver. Let $U_t \in \mathcal{U} = \{0,1\}$ denote the sensor's decision variable, where $U_t = 1$ indicates that a transmission is made at time $t$, while $U_t = 0$ indicates no transmission. We consider a packet-drop wireless communication channel. Let $H_t$ denote the packet-drop indicator, where $H_t = 1$ means that the packet transmitted at time $t$ is successfully decoded at the receiver, and $H_t = 0$ indicates an erasure event. The sequence $\{H_t\}$ is modeled as an i.i.d. Bernoulli process with
\begin{align}
    \Pr[H_t = 1] = p_s, ~
    \Pr[H_t = 0] = 1 - p_s = p_f.
\end{align}

\subsubsection{Receiver} Upon successful reception, the receiver updates its estimate using the newly received measurement\footnote{This is the so-called zero-order hold (ZOH) estimator. Better alternatives, such as the MAP estimator, can be found in~\cite{luo2025role, krishnamurthy2016POMDP}.}, i.e., $\hat{X}_t = X_{t}$, and sends an acknowledgment (ACK) packet to the sensor. Otherwise, a negative acknowledgment (NACK) is sent, and the remote estimate remains unchanged, i.e., $\hat{X}_t= \hat{X}_{t-1}$. We assume that ACK/NACK packets are delivered instantaneously and error-free.

The information available at the sensor at decision epoch $t$ is 
\begin{align}
    I_t = (X_{1:t}, \hat{X}_{1:t-1}, U_{1:t-1}).
\end{align}
At each time $t$, a decision $U_t$ is taken according to a transmission rule $\pi_t$, where
\begin{align}
    U_t = \pi_t(I_t) = \pi_t(X_{1:t}, \hat{X}_{1:t-1}, U_{1:t-1}).\label{eq:trans-policy}
\end{align}
A transmission policy is a sequence of transmission rules, i.e., $\pi=(\pi_1, \pi_2, \ldots)$. We call a policy stationary if it employs the same rule at every decision epoch. A policy is deterministic if, given the history $I_t$, it selects an action with certainty, while a randomized policy specifies a probability distribution on the action space $\mathcal{U}$. 

\subsubsection{Performance Measure} In classical systems, the system's performance is measured by \textit{average distortion}. Given a bounded distortion function
\begin{align}
    d_t : \mathcal{X}\times\mathcal{X}\rightarrow [0, \infty),\label{eq:distortion}
\end{align}
the average distortion of a policy $\pi$ is measured by
\begin{align}
    \mathcal{J}_\text{classic}(\pi) = \limsup_{T\rightarrow\infty}\mathbb{E}^{\pi} \Bigg[
    \frac{1}{T}\sum_{t=0}^{T-1}d_t(X_t, \hat{X}_t)
    \Bigg].
\end{align}
It is noteworthy that in classical systems, whether a measurement is discarded or transmitted does not depend on the significance of the measurement. As a result, distortion metrics are typically designed to penalize the system based solely on the physical discrepancy between the source signal $X_t$ and the reconstructed signal $\hat{X}_t$. A widely used distortion metric for Markov chains is the Hamming distortion, i.e.,
\begin{align}
    d_t(X_t, \hat{X}_t) = \mathds{1}\{X_t\neq \hat{X}_t\},\label{eq:classical-distortion}
\end{align}
where $\mathds{1}\{\cdot\}$ is the indicator function. 

In semantics-aware systems, however, information accuracy is not the only concern. The aim is to assess the value of information to reduce the amount of uninformative data transmissions. Let
\begin{align}
    \mathrm{VoI}_t = (\mathrm{VoI}^1_t, \mathrm{VoI}^2_t, \ldots, \mathrm{VoI}^K_t)
\end{align}
denote a set of semantic values conveyed by the current measurement $X_t$, where $\mathrm{VoI}^k_t$, $k = 1, 2, \ldots, K$, is the $k$th semantic attribute, extracted from the history of all system realizations up till time $t$, i.e.,
\begin{align}
    \mathrm{VoI}^k_t = \eta^k_t(X_{1:t}, \hat{X}_{1:t}, U_{1:t}),
\end{align}
where 
\begin{align}
    \eta^k_t: \mathcal{X}^{t} \times \mathcal{X}^{t} \times \mathcal{U}^{t} \rightarrow \Gamma^k
\end{align}
is an extraction function, and $\Gamma^k \subseteq [0, \infty)$ is the value domain. The performance of the semantics-aware system is measured by
\begin{align}
    \mathcal{J}(\pi) = \limsup_{T\rightarrow\infty}\mathbb{E}^{\pi} 
    \Bigg[
    \frac{1}{T}\sum_{t=0}^{T-1}c_t(X_t, \hat{X}_t, \mathrm{VoI}_t)
    \Bigg],\label{eq:semantics-aware cost}
\end{align}
where 
\begin{align}
    c_t: \mathcal{X} \times \mathcal{X} \times 
    \prod_{k=1}^{K}\Gamma^k \rightarrow [0, \infty)
\end{align}
is the cost function at time $t$, incorporating both the instantaneous estimation error $(X_t, \hat{X}_t)$ and its semantic value $\mathrm{VoI}_t$. Unlike the distortion function~\eqref{eq:distortion}, which depends solely on the current estimation error, the semantic attributes can be history-dependent. 

The goal is to minimize the average semantics-aware costs in \eqref{eq:semantics-aware cost} over constraints on available communications resources, such as channel bandwidth or energy budget. This problem is formulated as follows:
\begin{align}
    \inf_{\pi\in\Pi}\mathcal{L}^\lambda(\pi) = \mathcal{J}(\pi) +\lambda F(\pi) ,\label{problem:semantics-aware problem}
\end{align}
where $\Pi$ is the set of all admissible policies defined in \eqref{eq:trans-policy}, $\lambda$ is the cost associated with each transmission attempt, and
\begin{align}
    F(\pi) = \limsup_{T\rightarrow\infty}\mathbb{E}^{\pi} 
    \Bigg[
    \frac{1}{T}\sum_{t=0}^{T-1} \mathds{1}\{U_t\neq 0\}
    \Bigg]
\end{align}
is the transmission frequency under a policy $\pi$. When there is a hard constraint on the transmission frequency, the following constrained formulation is of interest: 
\begin{align}
    \inf_{\pi\in\Pi}\mathcal{J}(\pi),~\text{subject to}~ F(\pi) \leq F_{\max}, \label{problem:constrained-semantics-aware problem}
\end{align}
where $F_{\max} \in (0, 1]$ is the maximum allowed transmission frequency.

Problems~\eqref{problem:semantics-aware problem} and \eqref{problem:constrained-semantics-aware problem}, which will be discussed in greater detail in Section~\ref{sec:solution}, are unconstrained and constrained MDPs, respectively. These problems pose computational and memory challenges due to the (possibly) infinite state space and unbounded cost function. Fortunately, for the semantic attributes discussed later, we can construct a coherent body of theory, identify convenient structural properties of the optimal policy, and design efficient algorithms.

In the subsequent sections, we instantiate several key semantic attributes that have applications in many domains.

\subsection{Version Age of Information (VAoI)}
Incorporating \emph{versions} into semantics-aware communication systems has proven both beneficial and effective. The intuition is that, not all updates at the source provide new information or content that needs to be generated and transmitted to destination nodes; unnecessary updates may therefore waste system resources. Moreover, in large-scale networks -- where achieving coherent or synchronous timestamping across different nodes may be infeasible due to the lack of accurate, synchronized clocks\footnote{The investigation of AoI in the presence of clock drifts have been considered in  \cite{salimnejad2025relativistic}.} -- quantifying the AoI becomes unreliable.

\begin{figure}[t!]
    \centering
    \scalebox{0.67}{\begin{tikzpicture}[scale=1.0]
\def\dc{0.4}
\def\dt{0.8}
\def\T{16}
\def\Tm{15}
\draw[line width=0.5mm, ->] (0,0) -- (15.8*\dt,0) node[anchor=north] {\Large $t$};
\draw[line width=0.5mm, ->] (0,0) -- (0,9.8*\dc);

\foreach \i [evaluate=\i as \j using int(\i-1)] in {1,...,\T}{
    \fill (\j*\dt,0) circle[radius=1.5pt];
    \draw (\j*\dt,0) node[anchor=north] {$\i$}; 
}

\foreach \i in {1,...,\Tm}{
    \draw[dotted, gray, opacity=0.8, line width=0.2pt] (\i*\dt, 0) -- (\i*\dt, 9.5*\dc);
}
\foreach \i in {1,...,9}{
    \draw[dotted, gray, opacity=0.8, line width=0.2pt] (0,\i*\dc) -- (15.5*\dt,\i*\dc);
}

\draw[line width=0.5mm]  (0,0*\dc) to[bend right=0] (6*\dt,6*\dc); 
\draw[line width=0.5mm]  (6*\dt,6*\dc) to[bend right=0] (6*\dt,0*\dc); 
\draw[line width=0.5mm]  (6*\dt,0*\dc) to[bend right=0] (13*\dt,7*\dc); 
\draw[line width=0.5mm]  (13*\dt,7*\dc) to[bend right=0] (13*\dt,0*\dc); 
\draw[line width=0.5mm]  (13*\dt,0*\dc) to[bend right=0] (15*\dt,2*\dc); 

\draw[dashed, red, line width=0.5mm]  (0,0*\dc) to[bend right=0] (2*\dt,0*\dc);
\draw[dashed, red, line width=0.5mm]  (2*\dt,0*\dc) to[bend right=0] (3*\dt,1*\dc); 
\draw[dashed, red, line width=0.5mm]  (3*\dt,1*\dc) to[bend right=0] (6*\dt,1*\dc); 
\draw[dashed, red, line width=0.5mm]  (6*\dt,1*\dc) to[bend right=0] (6*\dt,0*\dc); 
\draw[dashed, red, line width=0.5mm]  (6*\dt,0*\dc) to[bend right=0] (8*\dt,0*\dc); 
\draw[dashed, red, line width=0.5mm]  (8*\dt,0*\dc) to[bend right=0] (10*\dt,2*\dc); 
\draw[dashed, red, line width=0.5mm]  (10*\dt,2*\dc) to[bend right=0] (11*\dt,2*\dc); 
\draw[dashed, red, line width=0.5mm]  (10*\dt,2*\dc) to[bend right=0] (11*\dt,2*\dc); 
\draw[dashed, red, line width=0.5mm]  (11*\dt,2*\dc) to[bend right=0] (12*\dt,3*\dc); 
\draw[dashed, red, line width=0.5mm]  (12*\dt,3*\dc) to[bend right=0] (13*\dt,3*\dc); 
\draw[dashed, red, line width=0.5mm]  (13*\dt,3*\dc) to[bend right=0] (13*\dt,0*\dc); 
\draw[dashed, red, line width=0.5mm]  (13*\dt,0*\dc) to[bend right=0] (15*\dt,0*\dc); 

\draw[->, line width=0.3mm]    (0,-0.5) -- (0,-0.9);
\draw[->, line width=0.3mm]    (6*\dt,-0.5) -- (6*\dt,-0.9);
\draw[->, line width=0.3mm]    (13*\dt,-0.5) -- (13*\dt,-0.9);

\draw[<-, line width=0.3mm]    (2*\dt,-0.5) -- (2*\dt,-0.9);
\draw[<-, line width=0.3mm]    (8*\dt,-0.5) -- (8*\dt,-0.9);
\draw[<-, line width=0.3mm]    (9*\dt,-0.5) -- (9*\dt,-0.9);
\draw[<-, line width=0.3mm]    (11*\dt,-0.5) -- (11*\dt,-0.9);

\begin{scope}[xshift=0.5cm, yshift=3.5cm]
    \draw[line width=0.5mm] (0,0) -- (0.8,0);
    \node[right] at (1.0,0) {\large AoI};

    \draw[dashed, red, line width=0.5mm] (0,-0.5) -- (0.8,-0.5);
    \node[right] at (1.0,-0.5) {\large VAoI};
\end{scope}

\end{tikzpicture}}
    \caption{Comparison of VAoI and AoI. Upward arrows ($\uparrow$) denote the generation of a new version at the source, whereas downward arrows ($\downarrow$) denote the reception of an update at the receiver.}
    \label{fig_VAoIAoI}
\end{figure}
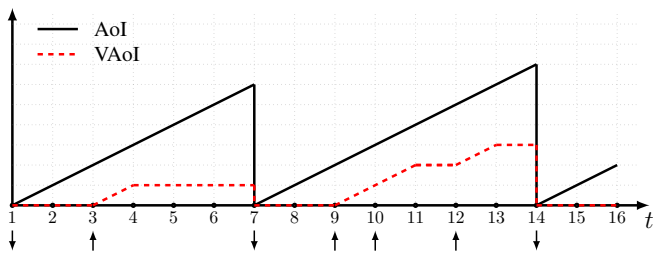

VAoI labels status updates with version numbers and measures how many versions out-of-date the information at a destination node is relative to the source~\cite{yates2021Vage}, i.e.,
\begin{align}
    \mathrm{VAoI}_t = N^\text{S}_t - N^\text{R}_t,
\end{align}
where $N^\text{S}_t$ and $N^\text{R}_t$ represent the version numbers of the current updates at the source and the receiver, respectively. More broadly, \emph{a new version can refer to any significant change in content or the generation of new information at the source}. From this perspective, VAoI serves as a semantic metric that captures both the freshness and relevance of information---without relying on timestamps---by measuring how many versions the receiver is lagging behind the source. Fig.~\ref{fig_VAoIAoI} illustrates the time evolution of both VAoI and AoI as functions of time. The AoI shows age growth that is not aligned with actual information relevance, resulting in unnecessary increases.

Table~\ref{table:VAoI_Eff} illustrates the performance of VAoI-optimal policies in a status update system with an energy-harvesting sensor~\cite{delfani2024qvaoi}. Incorporating the evolution of information versions significantly reduces the number of transmitted updates, leading to substantial gains in energy and cost efficiency. The VAoI-optimal policy achieves performance comparable to a greedy policy (which sends updates whenever energy is available) while reducing the number of transmissions by 54\%. In other words, to achieve this level of performance, the VAoI-optimal policy requires an energy-harvesting rate of less than $0.1$, whereas the greedy policy requires a rate of $0.2$.

\begin{table}[ht]
\centering
\caption{Comparison of the Greedy and VAoI-Optimal Policies}
\label{table:VAoI_Eff}
\begin{tabular}{c c c}
    \toprule
     & Greedy policy & VAoI-optimal policy \\
    \midrule
    Transmission freq. & 20\% & 9.27\% \\
    Avg. VAoI      & 1.49    & 1.48  \\
    \bottomrule
\end{tabular}
\end{table}

Most existing works on VAoI typically assume i.i.d. status updates, such as Poisson~\cite{yates2021age} and Bernoulli~\cite {delfani2024version} processes. The Version Innovation Age (VIA) proposed in~\cite{salimnejad2025age} extends VAoI to monitor Markovian sources. The VIA updates as follows:
\begin{align}
    \mathrm{VIA}_t = \begin{cases}
        \mathrm{VIA}_{t-1}, &X_t = X_{t-1}, U_tH_t = 0,\\
        \mathrm{VIA}_{t-1} + 1, &X_t \neq X_{t-1}, U_tH_t = 0,\\
        0, &U_tH_t = 1.
    \end{cases}
\end{align}

One limitation of the VIA metric is that it increases by one whenever a state change occurs and the transmission fails, even if the system remains synchronized. Age of Incorrect Versions (AoIV) addresses this limitation by measuring the number of outdated versions at the receiver when the system is in an erroneous state. AoIV is defined as~\cite{salimnejad2025age}
\begin{align}
    \mathrm{AoIV}_t = \begin{cases}
        \mathrm{AoIV}_{t-1}, &X_t = X_{t-1}, X_t \neq \hat{X}_t,\\
        \mathrm{AoIV}_{t-1} + 1, &X_t \neq X_{t-1}, X_t \neq \hat{X}_t,\\
        0, &X_t = \hat{X}_t.
    \end{cases} \label{Version_AoII}
\end{align}

\subsection{Content-Aware AoI}
An underlying assumption behind AoI is that information quality depends solely on its \emph{age}. However, information quality also depends on its \emph{content}\footnote{For instance, Shannon’s seminal work~\cite{shannon}, published in the 1940s, remains highly valuable.} and may evolve at different rates. Therefore, optimizing the system requires distinguishing not only \emph{when} information is generated, but also \emph{what} it contains.

\begin{figure}[t!]
    \centering
    \scalebox{0.67}{\begin{tikzpicture}[scale=1.0]
\def\dc{0.4}
\def\dt{0.8}
\def\T{16}
\def\Tm{15}
\draw[line width=0.5mm, ->] (0,0) -- (15.8*\dt,0) node[anchor=north] {\Large $t$};
\draw[line width=0.5mm, ->] (0,0) -- (0,9.8*\dc);

\foreach \i [evaluate=\i as \j using int(\i-1)] in {1,...,\T}{
    \fill (\j*\dt,0) circle[radius=1.5pt];
    \draw (\j*\dt,0) node[anchor=north] {$\i$}; 
}

\foreach \i in {1,...,\Tm}{
    \draw[dotted, gray, opacity=0.8, line width=0.2pt] (\i*\dt, 0) -- (\i*\dt, 9.5*\dc);
}
\foreach \i in {1,...,9}{
    \draw[dotted, gray, opacity=0.8, line width=0.2pt] (0,\i*\dc) -- (15.5*\dt,\i*\dc);
}

\draw (0.5*\dt,8.5*\dc) node{{ $\hspace{-0.3em}X\hspace{-0.3em}=\hspace{-0.2em}0$}};
\foreach \i [evaluate=\i as \j using {\i+0.5}] in {1,...,2}{
    \draw (\j*\dt,8.5*\dc) node{{$0$}};
}
\foreach \i [evaluate=\i as \j using {\i+0.5}] in {3,...,8}{
    \draw (\j*\dt,8.5*\dc) node{{$1$}};
}
\foreach \i [evaluate=\i as \j using {\i+0.5}] in {9,...,14}{
    \draw (\j*\dt,8.48*\dc) node{{$0$}};
}

\draw (0.5*\dt,7.5*\dc) node{{$\hat{X}\hspace{-0.3em}=\hspace{-0.2em}0$}};
\foreach \i [evaluate=\i as \j using {\i+0.5}] in {1,...,5}{
    \draw (\j*\dt,7.4*\dc) node{{$0$}};
}
\foreach \i [evaluate=\i as \j using {\i+0.5}] in {6,...,12}{
    \draw (\j*\dt,7.4*\dc) node{{$1$}};
}
\foreach \i [evaluate=\i as \j using {\i+0.5}] in {13,...,14}{
    \draw (\j*\dt,7.4*\dc) node{{$0$}};
}

\draw[line width=0.5mm]  (0,0*\dc) to[bend right=0] (6*\dt,6*\dc); 
\draw[line width=0.5mm]  (6*\dt,6*\dc) to[bend right=0] (6*\dt,0*\dc); 
\draw[line width=0.5mm]  (6*\dt,0*\dc) to[bend right=0] (9*\dt,0*\dc); 
\draw[line width=0.5mm]  (9*\dt,0*\dc) to[bend right=0] (13*\dt,4*\dc); 
\draw[line width=0.5mm]  (13*\dt,4*\dc) to[bend right=0] (13*\dt,0*\dc); 
\draw[line width=0.5mm]  (13*\dt,0*\dc) to[bend right=0] (15*\dt,2*\dc); 

\draw[dashed, red, line width=0.5mm]  (0,0*\dc) to[bend right=0] (3*\dt,0*\dc);
\draw[dashed, red, line width=0.5mm]  (3*\dt,0*\dc) to[bend right=0] (6*\dt,3*\dc); 
\draw[dashed, red, line width=0.5mm]  (6*\dt,3*\dc) to[bend right=0] (6*\dt,1*\dc); 
\draw[dashed, red, line width=0.5mm]  (6*\dt,0*\dc) to[bend right=0] (13*\dt,7*\dc); 
\draw[dashed, red, line width=0.5mm]  (13*\dt,7*\dc) to[bend right=0] (13*\dt,0*\dc); 
\draw[dashed, red, line width=0.5mm]  (13*\dt,0*\dc) to[bend right=0] (15*\dt,0*\dc); 

\draw[->, line width=0.3mm]    (0,-0.5) -- (0,-0.9);
\draw[->, line width=0.3mm]    (6*\dt,-0.5) -- (6*\dt,-0.9);
\draw[->, line width=0.3mm]    (13*\dt,-0.5) -- (13*\dt,-0.9);

\end{tikzpicture}}
    \caption{Evolution of the content-aware AoI~\eqref{eq:state-aware-AoI}. New measurements are received at time slots $t = 1$, $7$, and $14$. The red dashed line represents the AoI associated with the alarm state ($\Delta_t^1$), while the black solid line represents the AoI associated with the normal state ($\Delta_t^0$).}
    \label{fig:state-aware-AoI}
\end{figure}
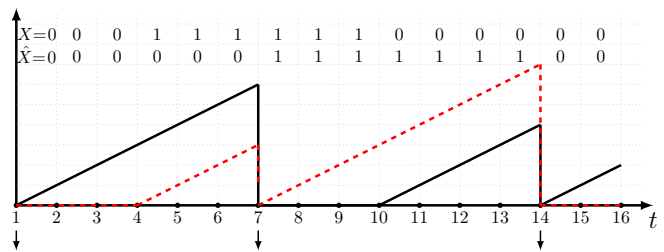

The content-aware AoI \cite{stamatakis2019control} explicitly captures this aspect by tracking the staleness of each state's information separately. Specifically, two distinct age variables, $\Delta_t^0$ and $\Delta_t^1$, are assigned to the normal and alarm states, respectively. The age associated with each state $x \in \{0, 1\}$ is recursively defined as
\begin{align}
    \Delta_t^x = \begin{cases}
        t - G^\text{R}_t, &\hat{X}_t=x,\\
        t - G^\text{C}_t, &\hat{X}_t\neq x, X_{t}=x,\\
        0, &\hat{X}_t \neq x,X_{t}\neq x,
    \end{cases}\label{eq:state-aware-AoI}
\end{align}
where $G^\text{R}_t$ is the timestamp of the most recent packet received at the receiver by time $t$, and $G^\text{C}_t$ is the time index of the most recent state change in the Markov source by time $t$. Fig.~\ref{fig:state-aware-AoI} depicts the evolution of content-aware AoI. The first case of~\eqref{eq:state-aware-AoI} tracks freshness based on the last received packet, which aligns with the definition of vanilla AoI. If the receiver's estimate remains $x$ despite the source temporarily leaving and returning to $x$, $\Delta_t^x$ continues to track the time since the last update, reflecting the receiver's unawareness of intermediate state changes. The second case tracks freshness from the most recent state change. For example, when the source changes its state from $0$ to $1$ while the receiver has not been notified about the change, $\Delta_t^0$ alone may not suffice, and another AoI variable $\Delta_t^1$ is activated as a complement to the vanilla AoI. In the third case, the AoI variable associated with state $x$ is inactive since neither the receiver nor the source references $x$. 

As shown in Fig.~\ref{fig:state-aware-AoI}, the AoI can be represented as the envelope of the content-aware AoI, i.e.,
\begin{align}
    \mathrm{AoI}_t  = \max\{\Delta_t^0, \Delta_t^1\}.
\end{align}
Since AoI treats all information updates equally, it may overestimate the importance of certain measurements. In contrast, content-aware AoI enables a more nuanced assessment of information staleness. For example, outdated updates from the alarm mode may lead to severe performance degradation, whereas aged updates from a normal mode may be more tolerable. The cost function can be written as~\cite{stamatakis2019control}
\begin{align}
    c(\Delta_t^0, \Delta_t^1) = f_0(\Delta_t^0) + f_1(\Delta_t^1),
\end{align}
where $f_x:\mathbb{N}\to[0,\infty), x\in\{0,1\}$, are increasing functions. For results on minimizing content-aware AoI with an energy-harvesting sensor, we refer the reader to~\cite{stamatakis2019control, delfani2025state}.

To better illustrate this metric, we re-examine the remote estimation example discussed in Section~\ref{sec:nonlinear-aging}. In practical NCSs, the plant, sensor, and actuator may not always operate in their desired modes. The system can shift to an abnormal state due to abrupt changes in system dynamics. Such changes may arise from anomalies, sudden environmental disturbances, component failures (e.g., a faulty sensor), or other unforeseen events. A formal treatment of such systems is the Markov jump linear system (MJLS)~\cite{costa2005mjls}, given by
\begin{align}
    Y_{t+1} = A_{X_t} Y_t + W_t,
\end{align}
where $\{Y_t\}$ is the plant process, and $\{X_t\}$ is the system mode process, which follows a time-homogeneous binary Markov chain. Here, $X_t=0$ denotes the pre-change (normal) mode, while $X_t=1$ represents the post-change (alarm) mode. The spectral radius of $A_1$ is greater than that of $A_0$, meaning that the system is more unstable in the alarm mode. Crucially, the evolution of the error covariance depends on the system mode. In the alarm mode, estimation errors may accumulate rapidly due to unstable dynamics, whereas in the normal mode, errors grow more slowly. This highlights that information quality is shaped not only by the age of the measurements but also by the latent content of the source. 

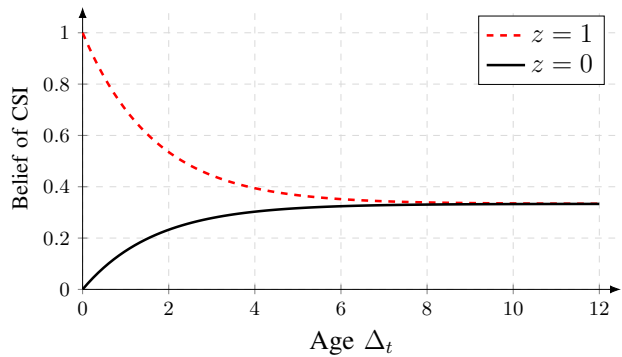
\begin{figure}[t!]
    \centering
    \scalebox{0.85}{\begin{tikzpicture}
    \begin{axis}[
        axis lines = left,
        xlabel = {\large Age $\Delta_t$},
        ylabel = {Belief of CSI}, 
        legend pos = north east,
        domain = 0:12,        
        samples = 100,
        xmin = 0, xmax = 12.5,
        ymin = 0, ymax = 1.1,
        width=10cm,           
        height=6cm,     
        grid = both,
        grid style={dashed,gray!30},  
        axis line style = {-Latex},   
        major grid style = {dashed,gray!30}, 
        tick label style={font=\small},  
        xtick = {0,2,...,12},
        ytick = {0,0.2,...,1.0},        
    ]
    \pgfmathsetmacro{\p}{0.15}
    \pgfmathsetmacro{\q}{0.3}
    
    \addplot[red, line width=0.4mm, dashed] {\p/(\p+\q) + \q/(\p+\q)*(1-\p-\q)^\x};
    \addlegendentry{\large $z = 1$}

    \addplot[black, line width=0.4mm] {\p/(\p+\q) - \p/(\p+\q)*(1-\p-\q)^\x};
    \addlegendentry{\large $z = 0$}
    \end{axis}
\end{tikzpicture}}
    \caption{The belief of the channel being in the \textsc{good} state vs. age of CSI.}
    \label{fig:AoCSI}
\end{figure}

Another example where content awareness is crucial in AoI would be the estimation of channel state information (CSI) \cite{lipski2024age, OnurPIMRC2025}. Consider a time-correlated fading channel $X_t$ with two states: a high reliability \textsc{good} state ($X_t = 1$) and a deep fading \textsc{bad} state ($X_t = 0$). The action $U_t = 1$ now stands for the decision to acquire the current CSI $X_t$, and $U_t = 0$ means using the previously observed channel state. The goal is to infer when the channel is in a \textsc{good} state for optimized data scheduling and resource allocation. Suppose the channel was last probed at time $G_t < t$ in state $z$, $z\in\{0, 1\}$. The probability (belief) of correctly identifying a \textsc{good} channel based on this outdated information is given by
\begin{align}
    \Pr[X_t &= 1 | \hat{X}_t = X_{G_t} = z] \notag\\
    &=\begin{cases}
        \frac{p}{p+q} - \frac{p}{p+q}(1-p-q)^{\Delta_t}, &z = 0,\\
        \frac{p}{p+q} + \frac{q}{p+q}(1-p-q)^{\Delta_t}, &z = 1.
    \end{cases}
\end{align}
Fig.~\ref{fig:AoCSI} illustrates the information quality as a function of the AoI $\Delta_t = t-G_t$ given outdated information $z$. We observe that the usefulness of aged CSI depends on when it is acquired and what it contains.

\subsection{Cost of Actuation Error (CAE)}
A key assumption behind traditional distortion measures is that the cost of estimation error depends solely on the physical discrepancy between the source state $X_t$ and the reconstructed signal $\hat{X}_t$. However, as illustrated by the autonomous driving example, different estimation errors can have different repercussions on system performance due to erroneous actions.

The CAE metric captures the fact that the cost of estimation error depends not only on the physical discrepancy but also on the contextual relevance and potential control risks to system performance~\cite{pappas2021goal}. It is defined as
\begin{align}
    \bar{d}_t(X_t, \hat{X}_t) = D_{X_t, \hat{X}_t}\mathds{1}\{X_t\neq \hat{X}_t\},\label{eq:context-aware distortion}
\end{align}
where $D_{X_t, \hat{X}_t}$ represents the potential control cost associated with error $(X_t, \hat{X}_t)$. In our problem, the alarm state (labeled as state $1$) is of greater importance. Intuitively, missed alarms typically incur higher costs than false alarms at the point of actuation. Performance analysis of this metric was presented in~\cite{pappas2021goal, salimnejad2023state, salimnejad2024real}. Optimal policies for minimizing CAE can be found in~\cite{fountoulakis2023goal, zakeri2024semantic, luo2024goal, luo2025semantic}. In particular, \cite{luo2024goal, luo2025semantic} developed optimal and learning-based communication policies in resource-constrained multi-source systems, a setting relevant for multimodal scenarios.

This metric is useful for achieving collaborative goals in distributed systems~\cite{luo2025semantic}. Consider, for example, a collaborative beamforming system consisting of two transmitters, $i = 1, 2$, and a single receiver. Each transmitter compensates for its channel phase offsets to ensure coherent signal addition at the receiver, thereby improving the overall beamforming gain. The problem is formalized as follows~\cite{lipski2024age}: The wireless link between each transmitter $i$ and the receiver is modeled as a complex reciprocal channel, $h^i_t = \alpha_i e^{jX_t^i}$, where $\alpha_i$ represents the path loss and $X_t^i$ denotes the phase offset introduced by the channel at time $t$. The receiver is assumed to be in the far field of the transmitter. Thus, $\alpha_i$ can be simple free-space path loss coefficients. The phase offset process $\{X_t^i\}_{t\geq 1}, X_t^i\in\{\beta_i^0, \beta_i^1\}$, is modeled by a binary Markov chain. The beamforming gain at the receiver is obtained as $\big|\sum_{i=1}^2 \alpha_i e^{j(X_t^i + \phi^i_t)}\big|^2$, where $\phi^i_t$ is the phase of the baseband signal at transmitter $i$. In collaborative beamforming, each transmitter $i$ adjusts its baseband phase $\phi^i_t$ to compensate for the channel phase offset $X_t^i$, such that the received signals from different transmitters add up coherently at the receiver. Specifically, if transmitter $i$ estimates the phase offset as $\hat{X}^i_t$, it applies a phase correction of $\phi^i_t =  - \hat{X}^i_t$. Consequently, the beamforming gain at the receiver becomes $\big|\sum_{i=1}^2 \alpha_i e^{j(X_t^i-\hat{X}^i_t)}\big|^2$. Let $X_t=(X^1_t, X^2_t)$ and $\hat{X}_t=(\hat{X}^1_t, \hat{X}^2_t)$. The cost function in question is
\begin{align}
    D_{X_t, \hat{X}_t} = (\alpha_1 + \alpha_2)^2 -  \Big|\sum_{i=1}^2 \alpha_i e^{j(X_t^i-\hat{X}^i_t)}\Big|^2,
\end{align}
which is a coupled cost that cannot be decomposed or proportionally assigned to individual transmitters. 

Table~\ref{tab:D_values} provides an example of the costs. From the table, incorrectly estimating the true channel states $X_t = (0^\circ, 120^\circ)$ as $\hat{X}_t = (60^\circ, 45^\circ)$ incurs a significantly higher cost than using the estimate $\hat{X}_t = (60^\circ, 120^\circ)$ when the channels are actually in state $X_t = (0^\circ, 45^\circ)$. When CSI is costly to acquire, it is more efficient to correct critical errors rather than treating all errors equally. This highlights the effectiveness of exploiting data significance in such systems. 
\begin{table}[h]
    \centering
    \renewcommand{\arraystretch}{1.2} 
    \setlength{\tabcolsep}{3pt} 
    \caption{Costs when $\alpha_1 = 2, \alpha_2 = 0.5, X^1_t\in\{0^\circ, 60^\circ\},X^2_t\in\{45^\circ, 120^\circ\}$.}
    \vspace{-0.1in}
    \begin{tabular}{|c|cccc|}
        \hline
        \diagbox[width=6em, height=2em, innerleftsep=4pt, innerrightsep=4pt]{\hspace{1em}\vspace{-0.1em}$X_t$}{\vspace{-2.5em} $\hat{X}_t$} & $(0^\circ, 45^\circ)$ & $(0^\circ, 120^\circ)$ & $(60^\circ, 45^\circ)$ & $(60^\circ, 120^\circ)$ \\
        \hline
        $(0^\circ, 45^\circ)$  & 0 & 1.48 & 1.0 & 0.07 \\
        $(0^\circ, 120^\circ)$ & 1.48 & 0 & 3.41 & 1.0 \\
        $(60^\circ, 45^\circ)$ & 1.0 & 3.41 & 0 & 1.48 \\
        $(60^\circ, 120^\circ)$ & 0.07 & 1.0 & 1.48 & 0 \\
        \hline
    \end{tabular}
    \label{tab:D_values}
\end{table}

\subsection{Cost of Consecutive Error (Lasting Impact)}
Another notable shortcoming of distortion measures lies in their \textit{history-independence}. Even though the source evolution is Markovian, the value of information depends on the history of past observations and decisions. For instance, as seen from the autonomous driving example, the longer an error persists, the more severe its consequences can become. 

The Age of Incorrect Information (AoII)~\cite{maatouk2020age} and the cost of memory error~\cite{salimnejad2024real} capture this notion by measuring the duration for which the system remains erroneous. The persistence cost is defined as
\begin{align}
    \mathrm{AoII}_t = d(X_t, \hat{X}_t)g(t - V_t),
\end{align}
where $d$ is a classical distortion measure, $g$ is a increasing function, $V_t$ denotes the last time the system was synced. In the literature, most existing works on AoII consider the Hamming distortion and a linear function $g$. Then, AoII can be recursively defined as
\begin{align}
    \mathrm{AoII}_t = \begin{cases}
        \mathrm{AoII}_{t-1} + 1, &X_t \neq \hat{X}_t,\\
        0, &X_t = \hat{X}_t.
    \end{cases}\label{eq:AoII}
\end{align}
However, this may not suffice since all errors are treated equally, i.e., context-agnostic. \emph{This egalitarianism can result in inadequate transmissions in urgent states but excessive transmissions in normal states.} 

In~\cite{luo2024minimizing, luo2024exploiting}, two new age metrics, the Age of Missed Alarm (AoMA) and the Age of False Alarm (AoFA), were introduced to measure the lasting impact of missed and false alarms, respectively. These age processes evolve as follows:
\begin{align}
    \Delta^\mathrm{AoMA}_t &= \begin{cases}
        \Delta^\mathrm{AoMA}_{t-1}, &X_t = 1, \hat{X}_t = 0,\\
        0, &\text{otherwise}.
    \end{cases}\\
    \Delta^\mathrm{AoFA}_t &= \begin{cases}
        \Delta^\mathrm{AoFA}_{t-1}, &X_t = 0, \hat{X}_t = 1,\\
        0, &\text{otherwise}.
    \end{cases}
\end{align}
Since at most one of these processes is active at any given time $t$, they can be expressed compactly as
\begin{align}
    \Delta_t & = \mathds{1}\{(X_t, \hat{X}_t) = (1, 0)\}\Delta^\mathrm{AoMA}_t \notag\\
        &\quad + \mathds{1}\{(X_t, \hat{X}_t) = (0, 1)\}\Delta^\mathrm{AoFA}_t.\label{eq:content-aware-age-metric}
\end{align}
The context-aware persistence cost is defined as
\begin{align}
    c_t(X_t, \hat{X}_t, \Delta_t) = \bar{d}(X_t, \hat{X}_t)\Delta_t,
\end{align}
where $\bar{d}$ is a context-aware distortion measure defined in~\eqref{eq:context-aware distortion}. 

\begin{figure}[t!]
    \centering
    \includegraphics[width=0.9\linewidth]{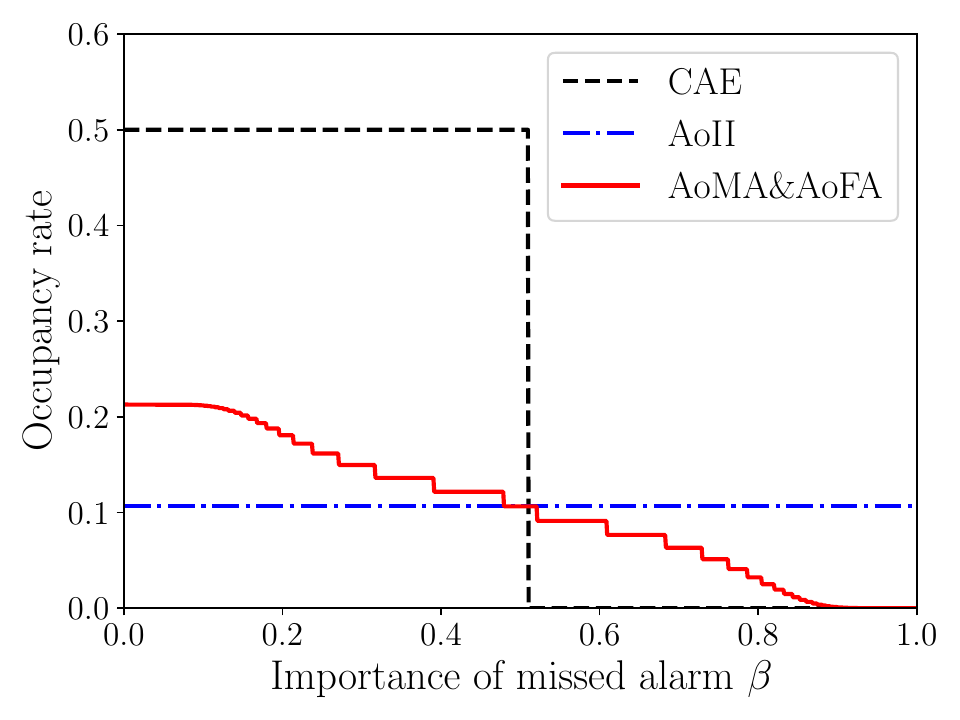}
    \caption{The occupancy rate of missed alarms when $p=q=0.25$ \cite{luo2024exploiting}.}
    \label{fig:occupancy-rate}
\end{figure}

Fig.~\ref{fig:occupancy-rate} compares the occupancy rates of missed alarms under different policies, where $\beta = D_{1,0}/(D_{0,1} + D_{1,0})$ is the relative importance of the missed alarms. A higher occupancy rate indicates that the system spends more time in missed alarms. The AoII-optimal policy does not adjust its behavior for different values of $\beta$. This is because AoII treats missed and false alarms equally, disregarding their relative significance on system performance. The CAE-optimal policy, on the other hand, triggers transmission only when an urgent error occurs~\cite{luo2025semantic}. When missed alarms incur relatively high (low) costs, transmission is initiated exclusively during missed (false) alarms, while false (missed) alarms are completely ignored. This results in significant lasting costs due to persistent unaddressed errors. In contrast, the AoMA\&AoFA-optimal policy dynamically adjusts its behavior based on the significance of missed alarms. As $\beta$ increases, the sensor transmits more frequently during missed alarms, accounting for a decrease in the occupancy rate.  

\subsection{Urgency of Lasting Impact}\label{sec:urgency-of-lasting-impact}
This section concerns the third semantic attribute we drew from the autonomous driving example: the lasting impact of an error depends on both its holding time and contextual significance. Motivated by this, \cite{luo2025cost} introduced the significance-aware Age of Consecutive Error (AoCE) to capture the urgency of lasting impact. The AoCE (i.e., error holding time) is defined as
\begin{align}
    \Delta_t = 
    \begin{cases}
        \Delta_{t-1}+ 1, &X_{t}\neq\hat{X}_{t}, (X_{t},\hat{X}_t)=(X_{t-1},\hat{X}_{t-1}),\\
        1, &X_t\neq\hat{X}_t,(X_{t},\hat{X}_t)\neq(X_{t-1},\hat{X}_{t-1}),\\
        0, &X_t=\hat{X}_t,
    \end{cases}\label{eq:aoce}
\end{align}
which extends AoMA-AoFA to general multi-state source models. The first case in \eqref{eq:aoce} applies when the error $(X_t, \hat{X}_t)$ persists. This occurs when the source remains unchanged and no packet is received at time $t$. The second case applies when the system enters a new error state, either because (i) no update is received at time $t$ while the source changes its state, or (ii) an update is received but the source transitions to a different state. In such cases, we reset $\Delta_t = 1$ to indicate the start of a new error episode. This differs from AoII \eqref{eq:AoII}, which continues to grow with error variations. In many control systems, such as autonomous driving, control costs or risks change as the error evolves. The system process $\{(X_t, \hat{X}_t, \Delta_t)\}_{t\geq 1}$ can be interpreted as a collection of interdependent age processes, each corresponding to a distinct estimation error. This structure allows AoCE to capture the lasting impact of different types of errors separately.

However, age alone may not suffice, as it ignores the significance (urgency) of the current estimation error. The significance-aware AoCE thus measures the urgency of lasting impact, i.e.,
\begin{align}
    c_t(X_t, \hat{X}_t, \Delta_t) = \bar{d}(X_t, \hat{X}_t)\cdot g_{X_t, \hat{X}_t}(\Delta_t),\label{eq:cost-function}
\end{align}
where $g_{i,j}(\cdot), i,j\in\set{X}$ are non-negative, non-decreasing, and possibly unbounded age functions. Here, $g_{i,j}(\delta)$ represents the cost of being in error $(i,j)$ for $\delta$ consecutive time slots. These age functions are quite general and may be discontinuous and non-convex.

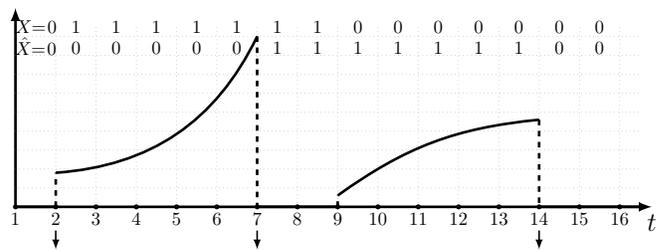
\begin{figure}[t!]
    \centering
    \scalebox{0.7}{\begin{tikzpicture}[scale=0.9]
\def\dc{0.4}
\def\dt{0.85}
\def\T{16}
\def\Tm{15}
\draw[line width=0.5mm, ->] (0,0) -- (15.8*\dt,0) node[anchor=north] {\Large $t$};
\draw[line width=0.5mm, ->] (0,0) -- (0,10.5*\dc);

\foreach \i [evaluate=\i as \j using int(\i-1)] in {1,...,\T}{
    \fill (\j*\dt,0) circle[radius=1.5pt];
    \draw (\j*\dt,0) node[anchor=north] {$\i$}; 
}

\foreach \i in {1,...,\Tm}{
    \draw[dotted, gray, opacity=0.8, line width=0.2pt] (\i*\dt, 0) -- (\i*\dt, 9.5*\dc);
}
\foreach \i in {1,...,9}{
    \draw[dotted, gray, opacity=0.8, line width=0.2pt] (0,\i*\dc) -- (15.5*\dt,\i*\dc);
}

\draw (0.5*\dt,9.5*\dc) node{{ $\hspace{-0.3em}X\hspace{-0.3em}=\hspace{-0.2em}0$}};
\foreach \i [evaluate=\i as \j using {\i+0.5}] in {1,...,7}{
    \draw (\j*\dt,9.5*\dc) node{{$1$}};
}
\foreach \i [evaluate=\i as \j using {\i+0.5}] in {8,...,14}{
    \draw (\j*\dt,9.5*\dc) node{{$0$}};
}

\draw (0.5*\dt,8.5*\dc) node{{$\hat{X}\hspace{-0.3em}=\hspace{-0.2em}0$}};
\foreach \i [evaluate=\i as \j using {\i+0.5}] in {1,...,5}{
    \draw (\j*\dt,8.4*\dc) node{{$0$}};
}
\foreach \i [evaluate=\i as \j using {\i+0.5}] in {6,...,12}{
    \draw (\j*\dt,8.4*\dc) node{{$1$}};
}
\foreach \i [evaluate=\i as \j using {\i+0.5}] in {13,...,14}{
    \draw (\j*\dt,8.4*\dc) node{{$0$}};
}

\draw[line width=0.5mm]  (0,0*\dc) to[bend right=0] (1*\dt,0*\dc);
\draw[dashed, line width=0.5mm]  (1*\dt,0*\dc) to[bend right=0] (1*\dt,1.8*\dc);
\draw[line width=0.5mm]  (1*\dt,1.8*\dc) to[bend right=30] (6*\dt,9*\dc);
\draw[dashed, line width=0.5mm]  (6*\dt,9*\dc) to[bend right=0] (6*\dt,0*\dc); 
\draw[line width=0.5mm]  (6*\dt,0*\dc) to[bend right=0] (8*\dt,0*\dc); 
\draw[dashed, line width=0.5mm]  (8*\dt,0*\dc) to[bend right=0] (8*\dt,0.6*\dc); 
\draw[line width=0.5mm]  (8*\dt,0.6*\dc) to[bend left=15] (13*\dt,4.6*\dc);  
\draw[dashed, line width=0.5mm]  (13*\dt,4.6*\dc) to[bend left=0] (13*\dt,0*\dc);  
\draw[line width=0.5mm]  (13*\dt,0*\dc) to[bend left=0] (15.5*\dt,0*\dc);  

\draw[->, line width=0.3mm]    (1*\dt,-0.5) -- (1*\dt,-0.9);
\draw[->, line width=0.3mm]    (6*\dt,-0.5) -- (6*\dt,-0.9);
\draw[->, line width=0.3mm]    (13*\dt,-0.5) -- (13*\dt,-0.9);

\end{tikzpicture}}
    \caption{Evolution of the significance-aware AoCE~\eqref{eq:aoce}. New measurements arrive at time slots $t=2$, $7$, and $14$. Estimation errors occur at $t=2$ and $t=9$ due to state changes in the source. Exponential and logarithmic functions are assigned to missed and false alarms, respectively.}
    \label{fig:nonlinear-AoCE}
\end{figure}

Fig.~\ref{fig:nonlinear-AoCE} illustrates the evolution of the significance-aware AoCE metric. Specifically, the context-aware distortion $\bar{d}$, AoCE $\Delta_t$, and non-linear age functions $g_{i,j}$ represent, respectively, the significance of the current estimation error, its lasting impact, and the urgency of lasting impact. For instance, in high-risk driving scenarios, errors may become increasingly critical over time, necessitating higher costs and exponential age functions to reflect the escalating risk. Conversely, for less critical errors, logarithmic or bounded age functions may suffice to represent their lasting impact, which grows at a diminishing rate.

\section{Optimizing the Goal-Oriented System Using Semantics-Aware Metrics}\label{sec:solution}
In this section, we present key analytical tools to characterize the existence and structure of optimal transmission policies, thereby addressing the value-of-timing problem in semantics-aware communication systems. Unlike the value of information, which is extracted from \emph{historical} data, the value of timing leverages predicted \emph{future} information to determine the optimal moment to initiate transmission. For example, even in the presence of critical errors, it may be preferable to delay transmission if the system is likely to return to normal conditions within a few time slots.

Based on how much future information is used in planning, transmission policies can be categorized as follows:
\begin{itemize}
    \item \textit{Myopic Policies:} These rely solely on the current state, without incorporating any predictions or foresight about future realizations. Randomized, reactive, and periodic policies fall into this category. While simple to implement, myopic policies often result in suboptimal transmission timing.
    \item \textit{Finite-Step Lookahead:} These policies predict possible trajectories over multiple time steps. Examples include Lyapunov optimization~\cite{neely2010stochastic}, which uses one-step predictions to balance constraint satisfaction and cost minimization, and Model Predictive Control (MPC)~\cite{MPC}, a widely used approach in control and robotics.
    \item \textit{Full-Horizon Planning:} These policies account for all possible future trajectories until the end of the horizon (in our setup, infinite), yielding global optimal transmission timing. Markov decision theory serves as the primary analytical tool in this category~\cite{puterman1994markov}.
\end{itemize}

In the following, we focus primarily on the MDP framework and present results for the significance-aware AoCE metric introduced in Section~\ref{sec:urgency-of-lasting-impact}. This metric represents the most challenging case; by developing solutions for it, we establish a framework that can be applied to other metrics as well. We first address the unconstrained Problem~\eqref{problem:semantics-aware problem}, followed by approaches to the constrained Problem~\eqref{problem:constrained-semantics-aware problem}. We will also cover some myopic policies and the Lyapunov approach.

\subsection{Unconstrained Formulation}\label{sec:unconstrained-formulation}
This section addresses the unconstrained MDP in~\eqref{problem:semantics-aware problem}. We first recap some preliminaries of MDPs. In optimal decision-making, it is of primary importance to identify an \emph{information state} $S_t \subseteq I_t$ to constrain the policy search space. An information state is a minimal sufficient statistic that summarizes all relevant information for decision-making; once it is identified, all other historical records can be discarded without losing optimality. To minimize AoCE, it suffices to maintain the following information state~\cite{luo2025cost}:
\begin{align}
    S_t = (X_t, \hat{X}_t, \Delta_t),
\end{align}
where $\Delta_t$ is the AoCE defined in~\eqref{eq:aoce}. Different information states may be used for different metrics. For example, distortion depends only on the current estimation error $(X_t, \hat{X}_t)$, while AoI relies solely on the information age.

The MDP in question is characterized by the tuple $(\mathcal{S}, \mathcal{U}, P, l)$. Herein, $\mathcal{S} = \mathcal{X}\times\mathcal{X}\times \mathbb{N}$, $\mathcal{U} = \{0, 1\}$, and $P(S_{t+1}|S_t, U_t)$ are the state space, action space, and transition kernel of the MDP, respectively. $l(S_t, U_t) = c(S_t) + \lambda U_t$ is the immediate cost of taking an action $U_t$ in state $S_t$, and $c(S_t)$ is the persistence cost at time $t$, defined in~\eqref{eq:cost-function}.

Referred to as the \emph{value (of timing) function}, $V(s)$ denotes the minimum expected total remaining cost when the current state is $s = (i, j, \delta)$. Let $V_\text{I}(s)$ and $V_\text{T}(s)$ denote, respectively, the expected total remaining cost when the sensor takes the idle action and the transmit action at the current time and then follows the optimal policy in the future. We then have
\begin{align}
    V(s) = \min\{V_\text{I}(s), V_\text{T}(s)\},\label{eq:value-function}
\end{align}
and
\begin{align}
    V_\text{I}(s) &= c(s) + \sum_{s^\prime\in\mathcal{S}}P(s^\prime|s,u=0)V(s^\prime),\label{eq:value-func-I}\\
    V_\text{T}(s) &= c(s) + \lambda + \sum_{s^\prime\in\mathcal{S}}P(s^\prime|s,u=1)V(s^\prime).\label{eq:value-func-T}
\end{align}
From~\eqref{eq:value-func-I} and~\eqref{eq:value-func-T}, the optimal transmission timing depends on both the immediate cost $l(s,u)$ and the (infinite-horizon) future expected costs, i.e., $\sum_{s^\prime}P(s^\prime|s,u)V(s^\prime)$. 

\subsubsection{Existence Result} Recall that AoCE can grow indefinitely, and the age functions $g_{i,j}$ are (possibly) unbounded. Consequently, the value function $V(s)$ may become unbounded regardless of the choice of policy~$\pi$. We are therefore interested in the conditions under which an optimal policy exists to achieve bounded average costs. Intuitively, the always-transmit policy provides the best estimation performance, albeit at the expense of the highest communication cost. Hence, a sufficient condition is that the estimation performance under the always-transmit policy $\tilde{\pi}$ is bounded, i.e., $\mathcal{J}(\tilde{\pi})<\infty$. As shown in~\cite{luo2025cost}, for an optimal policy to exist, the age functions, source pattern, and channel reliability must satisfy the following asymptotic growth condition:
\begin{align}
    \lim_{\delta\rightarrow\infty}\frac{g_{i,j}(\delta+1)}{g_{i,j}(\delta)}< \frac{1}{Q_{i, i}p_f}, \quad i\neq j, \label{eq:existence-condition}
\end{align}
where $Q_{i, i}p_f$ is the probability of remaining in error $(i,j)$ after each transmission attempt. Similar analyses can be applied to other metrics mentioned in Section~\ref{sec:semantics}. From~\eqref{eq:existence-condition}, we can conclude that an optimal policy trivially exists when the age functions are bounded, linear, or logarithmic. However, when exponential age functions are applied, the conditions in \eqref{eq:existence-condition} must be respected.

\subsubsection{Structural Results} We first define a partial order $\preceq$ on the state space $\mathcal{S}$. Specifically, for any  states 
$s_{1} = (i_1, j_1, \delta_1)$ and $s_{2} = (i_2, j_2, \delta_2)$, we define the ordering
\begin{align}
    s_1 \preceq s_2 \,\, \text{if} \,\, (i_1, j_1) = (i_2, j_2)\,\,\text{and}\,\,\delta_1 \leq \delta_2.
\end{align}
Note that states associated with different errors, i.e., $(i_1, j_1) \neq (i_2, j_2)$, are not comparable.

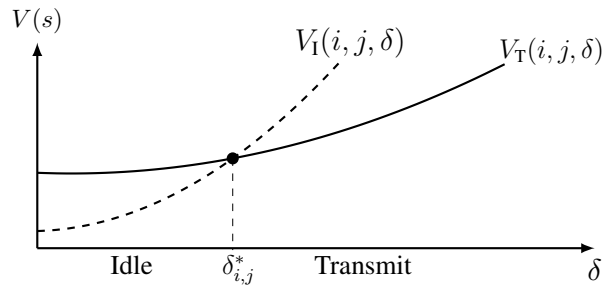
\begin{figure}[t!]
    \centering
    \scalebox{0.95}{\begin{tikzpicture}[scale=1.2]
    \draw[->, line width=0.3mm] (0, 0) -- (6.5, 0) node[below,scale=1.2] {$\delta$};
    \draw[->, line width=0.3mm] (0, 0) -- (0, 2.4) node[above,scale=1.] {$V(s)$};
    
    \node[scale=0.9] at (6.0, 2.3) {\large $V_\text{T}(i,j,\delta)$};
    \node at (3.65, 2.4) {\large $V_\text{I}(i,j,\delta)$};

    \draw[name path=A, domain=0.:3.55, smooth, variable=\x, dashed, line width=0.3mm] plot ({\x}, {0.15*(\x)*(\x+0.2)+0.2});

    \draw[name path=B, domain=0.:5.45, smooth, variable=\x, line width=0.3mm] plot ({\x}, {0.05*(\x+1.2)*(\x-2)+1});

    \path[name intersections={of=A and B, by=intersectionAB}];
    \fill[black] (intersectionAB) circle (2pt); 
    \draw[dashed] (intersectionAB) -- ++(0,-1.07);

    \node[scale=0.9] at (1.1, -.2) {\large $\textrm{Idle}$};
    \node[scale=0.9] at (2.35, -.25) {\large $\delta^*_{i,j}$};
    \node[scale=0.9] at (3.8, -.2) {\large $\textrm{Transmit}$};
\end{tikzpicture}}
    \caption{The monotonicity of the value function.}
    \label{fig:monotone-policy}
\end{figure}

Naturally, we expect the value function $V(s)$ and the optimal policy $\pi^*(s)$ to exhibit monotonicity properties; specifically, $V(s_1)\leq V(s_2)$ and $\pi^*(s_1) \leq \pi^*(s_1)$ for all $s_1 \preceq s_2$. Such structural properties are useful because they make it easy to search and implement the optimal policy. To formalize these results, we need to show that the transition kernel $P$ is stochastic monotone; that is, for each error $(i,j)$, the last term in \eqref{eq:value-func-I}-\eqref{eq:value-func-T}, i.e., $\sum_{s^\prime\in\mathcal{S}}P(s^\prime|s,u)V(s^\prime)$, is monotonically increasing in $\delta$. These results are established in~\cite{luo2025cost}, and we summarize the main findings here. For any fixed error, the monotonicity of the value functions implies that the sensor initiates transmission whenever the AoCE $\delta$ exceeds a fixed threshold $\tau_{i,j}^* \geq 1$, as illustrated in Fig.~\ref{fig:monotone-policy}. Therefore, the optimal policy exhibits a \emph{switching structure} that switches between different transmission thresholds depending on the error and its duration. Formally,
\begin{align}
    \pi^*(i,j,\delta) = \begin{cases} 
        1, &\delta \geq \tau_{i,j}^*,i\neq j,\\ 
        0, & {\text{otherwise}},
    \end{cases}\label{eq:switching-policy}
\end{align}
where $\tau_{i,j}^* = 1$ means always transmitting in estimation error $(i, j)$, whereas $\tau_{i,j}^* \to \infty$ means no transmission. 

\begin{figure}[t!]
    \centering
    \includegraphics[width=0.95\linewidth]{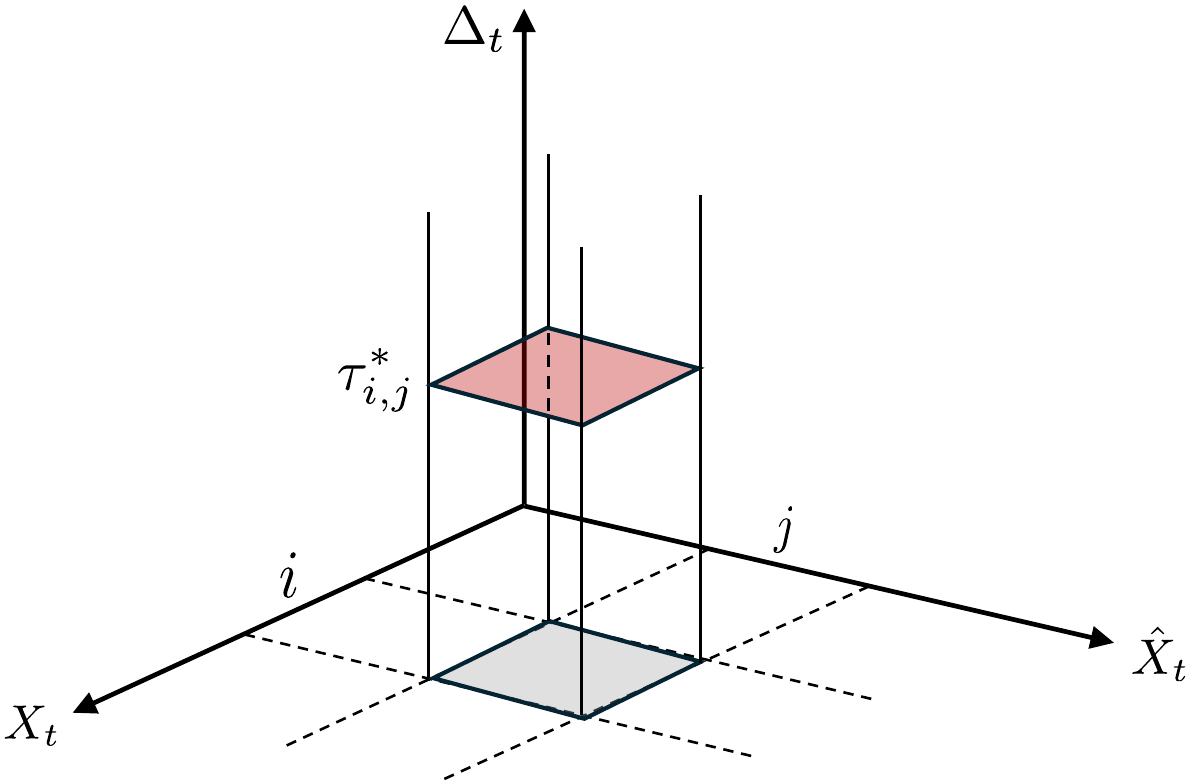}
    \caption{The structure of an optimal policy in the age-distortion space~\cite{luo2025cost}.}
    \label{fig:policy-structure}
\end{figure}

Fig.~\ref{fig:policy-structure} depicts the structure of the optimal policy in the age-distortion space. This reveals the value of timing in such systems. According to \cite{luo2025semantic}, distortion-optimal policies only tell ``whether to transmit when a certain error occurs". That is, the optimal threshold for the error $(X_t, \hat{X}_t) = (i,j)$ is either $\tau^*_{i,j} = 1$ (i.e., always transmit) or $\tau^*_{i,j} \rightarrow \infty$ (i.e., never transmit). By incorporating the error holding time as a third dimension in the decision-making process, we can further determine the \emph{optimal timing} to initiate transmission, allowing for transmissions to occur after several consecutive errors, i.e., $\delta_{i,j}^*\geq 1$. Moreover, this result circumvents the ``curse of memory" and the ``curse of dimensionality" of MDPs. One only needs to compute a small number of thresholds offline and store them in the sensor memory instead of solving a high-dimensional dynamic programming recursion and saving the results for infinitely many states. 

\subsubsection{Asymptotic Optimality} 
Although the switching structure~\eqref{eq:switching-policy} considerably reduces the policy searching space, finding the optimal thresholds is still challenging, as classic dynamic programming methods cannot iterate over infinitely many states~\cite{puterman1994markov}. For numerical tractability, we truncate the age process and propose a finite-state approximate MDP. The truncated AoCE is defined as
\begin{align}
    \Delta_{t}(N) = \min\{\Delta_t, N\}\label{eq:truncated-aoce},
\end{align}
where $N>0$ is a finite truncation bound. The truncated MDP is described by the tuple $(\bar{\mathcal{S}}, \mathcal{U}, \bar{P}, l)$, where $\bar{\mathcal{S}} = \mathcal{X}\times\mathcal{X}\times\{0, 1, \ldots, N\}$ is the truncated state space, and $\bar{P}$ is the transition kernel on the truncated space.

An important result established in~\cite{luo2025cost} shows that the truncated MDP converges exponentially fast to the original MDP in the truncation bound. Therefore, we shall feel safe to truncate the AoCE with an appropriately chosen bound.

Building on these findings, \cite{luo2025cost} proposed a Structured Policy Iteration (\texttt{SPI}) algorithm, which far outperforms the classic dynamic programming methods in terms of complexity. The \texttt{SPI} algorithm proceeds as follows:
\begin{enumerate}
    \item[a.] \textit{Initialization:} Arbitrarily select an initial policy $\pi^0$, a reference state $s_\textrm{ref}$. Choose a truncation bound $N$ such that $(Q_{i,i}p_f)^N < \epsilon$ for all $i\in\set{X}$, where $\epsilon>0$ is an arbitrarily small constant.
    \item[b.] \textit{Policy Evaluation:} Find a scalar $\set{L}^n$ and a vector $h^n$ by solving
    \begin{align}
        \set{L}^n + h^n(s) = l(s, \pi^n(s)) + \hspace{-0.2em}\sum_{s^\prime}P_{s,s^\prime}(\pi^n(s))h^n(s^\prime)\notag
    \end{align}
    for all $s\in \bar{\mathcal{S}}$ such that $h^n(s_\textrm{ref}) = 0$.
    \item[c.] \textit{Policy Improvement:} For each $(i, j)$, update $\pi^{n+1}(i, j, \delta)$ in the increasing order of the AoCE $\delta$:
    \begin{enumerate}
        \item[i.] Initialize $s = (i, j, \delta)$ with $\delta = 1$. 
        \item[ii.] Update the policy as
        \begin{align}
            \pi^{n+1}(s)
           \hspace{-0.2em} = \hspace{-0.2em} \argmin_{a\in\set{A}} \bigg[l(s, a) + \hspace{-0.5em}\sum_{s^\prime}\hspace{-0.5em}P_{s,s^\prime}(a)h^n(s^\prime)\bigg].\notag
        \end{align}
        \item[iii.] If $\pi^{n+1}(s) = 1$, the optimal action for all subsequent states $s^\prime \succeq s$ is to transmit without further computation. Thus, set $\pi^{n+1}(s^\prime) = 1$ for all $s^\prime \succeq s$ and proceed to Step (c) with an unvisited error. Otherwise, increment $\delta$ by setting $s = (i, j, \delta + 1)$ and return to Step~(ii).
    \end{enumerate}
    \item[d.] \textit{Stopping Criterion:} If $\pi^{n+1} = \pi^n$, the algorithm terminates with $\set{L}^* = \set{L}^n$ and $\pi^* = \pi^n$; otherwise increase $n=n+1$ and return to Step~(b).
\end{enumerate}

\subsubsection{Numerical Example} We now discuss some results reported in~\cite{luo2025cost}. Consider an asymmetric source with $|\mathcal{X}| = 4$ states and transition probability matrix 
\begin{align}
    Q = \begin{bmatrix}
        0.7 & 0.1 & 0.1 & 0.1\\
        0.05& 0.7 & 0.15& 0.1\\
        0.1 & 0.1 & 0.6 & 0.2\\
        0.05& 0.1 & 0.05& 0.8
    \end{bmatrix}.\notag
\end{align}
Assign exponential and logarithmic age functions to missed and false alarms, respectively, i.e.,
\begin{align}
    D_{i,j}g_{i,j}(\delta) = \begin{cases}
        e^{0.3\delta}, &i = 1, j\neq 1,\\
        \log(\delta) + 1, &i \neq 1, j=1,\\
        1, &\text{otherwise.}
    \end{cases}\notag
\end{align}

For comparison, we consider the following myopic policies:
\begin{itemize}
    \item \textit{Randomized}: The sensor transmits at every slot $t$ with a fixed probability $p_\alpha\leq 1$. 
    \item \textit{Periodic}: The sensor transmits every $t_\textrm{h}\geq 1$ slots, and remains salient otherwise.
    \item \textit{Reactive}: A new transmission is triggered only on state changes, i.e., $X_t \neq X_{t-1}$.
    \item \textit{Error-triggered}: A new transmission is triggered whenever an error occurs, i.e., $X_t \neq \hat{X}_{t-1}$.
\end{itemize}
Moreover, we also consider a \textit{threshold policy} which triggers transmission once the AoCE exceeds a given threshold $\delta_\textrm{th}\geq 1$, irrespective of the instantaneous estimation error. The optimal values of $p_\alpha^*$, $t_\textrm{h}^*$, and $\delta^*_\textrm{th}$ for the randomized, periodic, and threshold policies are obtained by brute force search. The optimal switching policy is obtained using the \texttt{SPI} algorithm. 

A performance comparison of these policies is presented in Table~\ref{table:performance-comparison}. When communication is cost-free ($\lambda=0$), nearly all policies reduce to the always-transmit policy. As $\lambda$ increases, the myopic policies become increasingly inefficient because they rely solely on the current estimation error and ignore the error holding time. The threshold policy significantly outperforms the myopic policies; however, it still exhibits a performance gap relative to the optimal switching policy due to its reliance on the insufficient statistic $\Delta_t$.

Table~\ref{table:optimal-policy-lambda} reports the optimal thresholds of various policies under different transmission costs $\lambda$. The results show that, when communication is costly or the channel quality is poor, the optimal switching policy transmits less frequently (or not at all) in false alarms and normal errors while consistently prioritizing missed alarms. In contrast, the distortion-optimal policy either triggers transmission for all errors (i.e., $\tau_{i,j} = 1, \forall i\neq j$) when communication is inexpensive ($\lambda\leq 1$) or remains silent otherwise (i.e., $\tau_{i,j} \rightarrow \infty, \forall i\neq j$). Due to source symmetry, the AoI- and AoII-optimal policies adopt a single threshold for all errors \cite{maatouk2020age}. The AoI metric shows obvious disadvantages as it completely ignores the source pattern and initiates transmissions even in synced states. The AoII metric is also inefficient in our problem since it treats all errors equally, resulting in excessive transmissions for less critical errors. Therefore, the significance-aware AoCE provides more informative decisions and offers a richer perspective than metrics based solely on distortion or information freshness.

\begin{table}[t!]
\centering
\scriptsize
\setlength{\tabcolsep}{4.6pt}
\caption{Performance comparison of different policies\cite{luo2025cost}}
\label{table:performance-comparison}
\renewcommand{\arraystretch}{0.8}
\begin{tabular}{cccccccc}
    \toprule
    \multirow{2.5}{*}{$\lambda$} & \multicolumn{6}{c}{Achievable minimal average costs} \\
    \cmidrule(lr){2-7}
    & Randomized & Periodic & Reactive & Error-triggered & Threshold & \textbf{Switching} \\
    \midrule
    0  & \textbf{0.31} & \textbf{0.31} & 0.43 & \textbf{0.31}  & \textbf{0.31}  & \textbf{0.31} \\
    1  & 0.93  &0.85   & 0.72 &0.62 & 0.62& \textbf{0.59} \\
    2  & 1.46  & 1.05  & 0.98 &0.91 & 0.88 & \textbf{0.73} \\
    3  & 1.20  & 1.20  & 1.27 &1.21 & 0.97 & \textbf{0.80} \\
    4  & 1.20  & 1.20  & 1.55 &1.50 & 1.01 & \textbf{0.85} \\
    5  & 1.20  & 1.20  & 1.83 &1.80 & 1.04 & \textbf{0.88} \\
    \bottomrule
\end{tabular}
\end{table}

\begin{table}[t!]
\centering
\scriptsize
\setlength{\tabcolsep}{3.3pt}
\caption{Optimal thresholds obtained by different policies \cite{luo2025cost}}
\label{table:optimal-policy-lambda}
\renewcommand{\arraystretch}{0.8}
\begin{tabular}{cccc|c|c|c}
    \toprule
    \multirow{2.5}{*}{$\lambda$} & \multicolumn{3}{c|}{Switching policy} & Distortion & AoI & AoII \\
    \cmidrule(lr){2-7} 
    & Missed alarms & False alarms & Normal errors & All errors & All states & All errors \\
    \midrule
    0   & 1 & 1       & 1     & 1       & 1       & 1 \\
    1   & 1 & 1       & 1       & 1       & 2       & 1 \\
    2   & 1 & 1       & $\infty$& $\infty$& 2       & 1 \\
    3   & 2 & 3       & $\infty$& $\infty$& 3       & 1 \\
    4   & 3 & 11      & $\infty$& $\infty$& 3       & 1 \\
    5   & 3 & $\infty$& $\infty$& $\infty$& 3       & 2 \\
    6   & 3 & $\infty$& $\infty$& $\infty$& 4       & 2 \\
    7   & 3 & $\infty$& $\infty$& $\infty$& 4       & 3 \\
    \bottomrule
\end{tabular}
\end{table}

\subsection{Constrained Formulation}
This section addresses the constrained MDP in~\eqref{problem:constrained-semantics-aware problem}. Three solution approaches are discussed: Lagrangian, Lyapunov, and token-based methods.

\subsubsection{Lagrangian Relaxation}
This approach relaxes the constraint by solving the following two-layer problem:
\begin{align}
    \sup_{\lambda\geq0} \inf_{\pi\in\Pi}\mathcal{L^\lambda}(\pi) - \lambda F_{\max},\label{problem:lagrangian-problem}
\end{align}
where $\lambda$ is the Lagrange multiplier, a tunable parameter rather than the fixed transmission cost used in the unconstrained Problem~\eqref{problem:semantics-aware problem}. The inner problem is equivalent to Problem~\eqref{problem:semantics-aware problem} since the term $\lambda F_{\max}$ is constant with respect to $\pi$. Let $\pi^*_\lambda$ denote the optimal policy to the inner problem for a fixed $\lambda$. For notational simplicity, we write $\mathcal{J}^\lambda = \mathcal{J}(\pi^*_\lambda)$, $F^\lambda = F(\pi^*_\lambda)$, and $\mathcal{L}^\lambda = \mathcal{J}^\lambda + \lambda F^\lambda$.

In general, there is no guarantee that a solution to the relaxed Problem~\eqref{problem:lagrangian-problem} is optimal to the original constrained problem. The following result (e.g., \cite{ross1985CMDP}) gives conditions for a solution to be optimal: If there exists a $\lambda>0$ such that $\mathcal{J}^\lambda<\infty$ and $F^\lambda = F_{\max}$, then $\pi^*_\lambda$ solves Problem~\eqref{problem:constrained-semantics-aware problem}. This result suggests a strategy to construct an optimal policy:
\begin{enumerate}
    \item[i.] If there exists a multiplier $\lambda > 0$ such that $F^\lambda = F_{\max}$, then the optimal policy is simply $\pi^* = \pi^*_\lambda$.
    \item[ii.] Suppose no such $\lambda$ as the above exists. Instead, find a multiplier $\lambda>0$ such that for an arbitrarily small constant $\epsilon$, it satisfies $F^{\lambda+\epsilon}<F_{\max}<F^{\lambda-\epsilon}$. The optimal policy is a \emph{randomized mixture} of two simple policies, denoted by $\pi^* = (p_\lambda, \pi^*_{\lambda-\epsilon}, \pi^*_{\lambda+\epsilon})$. That is, it selects policy $\pi^*_{\lambda-\epsilon}$ with probability $p_\lambda$ and policy $\pi^*_{\lambda+\epsilon}$ with probability $1-p_\lambda$ at each time step, where $p_\lambda$ satisfies $p_\lambda F^{\lambda-\epsilon} + (1-p_\lambda) F^{\lambda+\epsilon} = F_{\max}$. 
\end{enumerate}

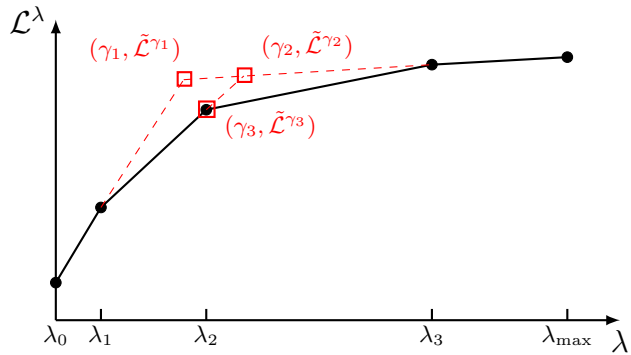
\begin{figure}[t!]
    \centering
    \scalebox{1}{\begin{tikzpicture}[every node/.style={font=\small}]
  \draw[->, thick] (0,0) -- (7.5,0) node[below] {\large $\lambda$};
  \draw[->, thick] (0,0) -- (0,4) node[left] {\large $\mathcal{L}^\lambda$};

  \node at (0,-0.2) {$\lambda_0$};
  \node at (0.6,-0.2) {$\lambda_1$};
  \draw[thick] (0.6,0.) -- (0.6, 0.15);
  \node at (2,-0.2) {$\lambda_2$};
  \draw[thick] (2,0.) -- (2, 0.15);
  \node at (5,-0.2) {$\lambda_3$};
  \draw[thick] (5,0.) -- (5, 0.15);
  \node at (6.8,-0.2) {$\lambda_{\max}$};
  \draw[thick] (6.8,0.) -- (6.8, 0.15);

  \coordinate (l0) at (0,0.5);
  \coordinate (l1) at (0.6,1.5);
  \coordinate (l2) at (2, 2.8);
  \coordinate (l3) at (5,3.4);
  \coordinate (lmax) at (6.8,3.5);

  \draw[thick] (l0) -- (l1) -- (l2) -- (l3) -- (lmax);

  \filldraw[black] (l0) circle (2pt);
  \filldraw[black] (l1) circle (2pt);
  \filldraw[black] (l2) circle (2pt);
  \filldraw[black] (l3) circle (2pt);
  \filldraw[black] (lmax) circle (2pt);

  \coordinate (g1) at (1.7,3.2);
  \coordinate (g2) at (2.5,3.25);

  \draw[red, thick] (g1) +(-0.08,-0.08) rectangle +(0.1,0.1) node[above left] {\small $(\gamma_1, \tilde{\mathcal{L}}^{\gamma_1})$};
  \draw[red, thick] (g2) +(-0.08,-0.08) rectangle +(0.1,0.1) node[above right] {\small $(\gamma_2, \tilde{\mathcal{L}}^{\gamma_2})$};
  \draw[red, thick] (l2) +(-0.096,-0.096) rectangle +(0.11,0.11) node[below right] {\small $(\gamma_3, \tilde{\mathcal{L}}^{\gamma_3})$};

  \draw[red, dashed] (l1) -- (g1);
  \draw[red, dashed] (g1) -- (l3);
  \draw[red, dashed] (l2) -- (g2);

  \coordinate (g1base) at (1.7,0);
  \coordinate (g2base) at (3.2,0);

\end{tikzpicture}}
    \caption{An illustration of the intersection search method \cite{luo2025semantic}. Consider that, within the interval $[0, \lambda_\textrm{max}]$, $\mathcal{L}^\lambda$ has $4$ segments and three corner points, i.e., $(\lambda_1, \mathcal{L}^{\lambda_1})$, $(\lambda_2, \mathcal{L}^{\lambda_2})$ and $(\lambda_3, \mathcal{L}^{\lambda_3})$. The squares represent the intersection points. The optimal multiplier is the second corner point, i.e., $\lambda^* = \lambda_2$.}
    \label{fig:intersection-search}
\end{figure}

Finding such a $\lambda$ value is computationally prohibitive. The intersection search (\texttt{Insec}) method~\cite{luo2025semantic} is an efficient multiplier update method that determines the optimal multiplier $\lambda^*$ in only a few iterations. This method exploits the following properties of the Lagrangian cost $\mathcal{L}^\lambda$: (1) $\mathcal{L}^\lambda$ is a piecewise linear, continuous, and increasing function of $\lambda$, and (2) the optimal multiplier $\lambda^*$ corresponds to a corner point of $\mathcal{L}^\lambda$, as illustrated in Fig.~\ref{fig:intersection-search}.

The \texttt{Insec} algorithm proceeds as follows:
\begin{enumerate}
    \item[a.] \textit{Initialization:} Choose a sufficiently large $\lambda_{\max}$ such that $F^{\lambda_{\max}} <F_{\max}$. Initialize $I^0 = [\lambda^0_{-},\lambda^0_{+}] = [0, \lambda_{\max}]$.
    \item[b.] \textit{Optimization:} Apply the \texttt{SPI} algorithm to solve the inner problem for two fixed multipliers, $\lambda_{-}^n$ and $\lambda_{+}^n$. 
    \item[c.] \textit{Intersection:} Compute the intersection point of two tangents: one is formed by point $(\lambda^n_{-}, \mathcal{L}^{\lambda^n_{-}})$ with a slop of $F^{\lambda^n_{-}}$, and another is formed by point $(\lambda^n_{+}, \mathcal{L}^{\lambda^n_{+}})$ with a slop of $F^{\lambda^n_{+}}$. The intersection point is given by $(\gamma^n, \tilde{\mathcal{L}}^{\gamma^n})$, where
    \begin{align}
        \gamma^n = \frac{\mathcal{J}^{\lambda_{+}^n} - \mathcal{J}^{\lambda_{-}^n}}{F^{\lambda_{-}^n} - F^{\lambda_{+}^n}}, ~ \tilde{\mathcal{L}}^{\gamma^n} = \mathcal{J}^{\lambda_{-}^n} + \gamma^n F^{\lambda_{-}^n}.
    \end{align}
    \item[d.] \textit{Multiplier update:} Update the search interval as: $I^n = [\gamma^n, \lambda^{n-1}_{+}]$ if $F^{\gamma^n}\geq F_{\max}$; otherwise, $I^n = [\lambda^{n-1}_{-}, \gamma^n]$.
    \item[e.] \textit{Stop criterion:} The algorithm terminates when the intersection point is located on $\mathcal{L}^\lambda$, i.e., $\tilde{\mathcal{L}}^{\gamma^n}=\mathcal{L}^{\gamma^n}$. Otherwise, set $n=n+1$ and go to Step (b).
\end{enumerate}

\subsubsection{Lyapunov Relaxation}
The Lyapunov approach is a finite-horizon method (more precisely, one-step lookahead) that converts the constrained MDP into a simple Min-Weight problem. At each step, it greedily balances \emph{queue stability} (i.e., constraint satisfaction) against cost minimization~\cite{neely2010stochastic}. 

\begin{figure}[t!]
    \centering
    \begin{subfigure}{\linewidth}
        \centering
        \scalebox{1}{\begin{tikzpicture}[>=stealth, node distance=1cm, every node/.style={align=center}]
    \draw[thick] (0,0) -- (3,0) -- (3,1) -- (0,1);
    \foreach \x in {1.5,2.0, 2.5}
        \draw[thick] (\x,0) -- (\x,1);

    \draw[thick] (3.5,0.5) circle (0.5);

    \draw[->, thick] (-1.5,0.5) -- (0,0.5) node[midway, above] {$U_t$};

    \draw[->, thick] (4,0.5) -- (5.5,0.5) node[midway, above] {$F_{\max}$};

    \node at (1.5,-0.3) {\small Queue};
    \node at (3.5,-0.3) {\small Channel};
\end{tikzpicture}}
        \caption{Lyapunov approach}
        \label{fig:virtual-queue-lyapunov}
        \end{subfigure}
    \\ \vspace{0.4em}
    \begin{subfigure}{\linewidth}
        \centering
        \scalebox{1}{\begin{tikzpicture}[>=stealth, node distance=1cm, every node/.style={align=center}]

    \draw[thick] (0,0) rectangle (3,1);
    \foreach \x in {1.5,2.0, 2.5}
        \draw[thick] (\x,0) -- (\x,1);

    \draw[thick] (3.5,0.5) circle (0.5);

    \draw[->, thick] (-1.5,0.5) -- (0,0.5) node[midway, above] {$F_{\max}$};

    \draw[->, thick] (4,0.5) -- (5.5,0.5) node[midway, above] {$U_t$};

    \node at (1.5,1.3) {{\small Buffer size:}~$\Omega_{\max}$};

    \node at (1.5,-0.3) {\small Queue};
    \node at (3.5,-0.3) {\small Channel};
\end{tikzpicture}}
        \caption{Token-based approach}
        \label{fig:virtual-queue-token}
    \end{subfigure}  
    \caption{Virtual queue representation of different approaches.}
    \label{fig:virtual-queues}
\end{figure}
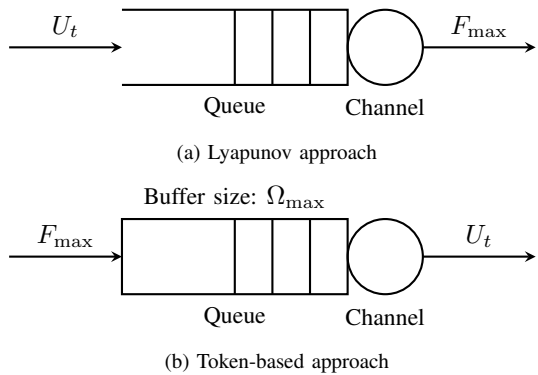

Let $\Theta_t$ denote a virtual queue associated with the transmission constraint in Problem~\eqref{problem:constrained-semantics-aware problem}, defined as
\begin{align}
    \Theta_{t+1} = \big[\Theta_t - \Xi_t\big]^{\infty}_{0} + U_t,\label{eq:virtual-queue}
\end{align}
where $\{\Xi_t\}_{t\geq 1}$ is an i.i.d. Bernoulli process with mean $F_{\max}$. The operation $\big[x\big]^{b}_{a} = \min\big\{\max\{x, a\},b\big\}$ bounds a variable $x$ between an interval $[a, b]$. Herein, $F_{\max}$ represents the virtual service rate of the queue $\Theta_t$, and $\{U_t\}_{t\geq 1}$ serves as a controllable arrival process, as depicted in~Fig.~\ref{fig:virtual-queue-lyapunov}. The Lyapunov theorem states that if the virtual queue $\Theta_t$ is \emph{mean rate stable}, i.e., $\lim_{t\rightarrow\infty}\mathbb{E}[\Theta_t]/t = 0$, the constraint in~\eqref{problem:constrained-semantics-aware problem} is satisfied with probability (w.p.) $1$ \cite[Ch.~4]{neely2010stochastic}. 

Let $\Gamma(\Theta_t) = \frac{1}{2}\Theta_t^2$ denote a scalar measure of queue congestion. At every time slot $t$, given current queue backlog $\Theta_t$ and system state $S_t$, the Lyapunov approach seeks to minimize the following drift-plus-penalty (\texttt{DPP}) expression:
\begin{align}
    \min_{U_t\in\mathcal{U}}
    \underbrace{\mathbb{E}
    \{\Gamma(\Theta_{t+1}) - \Gamma(\Theta_t)|\Theta_t\}}_{\text{drift}} + \zeta \underbrace{\mathbb{E}\{c(S_{t+1})|S_t, U_t\}}_{\text{penalty}}.\label{eq:dpp}
\end{align}
Here, the drift term measures the one-step expected change in queue backlog, while the penalty term represents the one-step expected estimation cost. The tuning parameter $\zeta > 0$ balances between queue backlog and cost minimization. More specifically, increasing $\zeta$ places more emphasis on improving estimation performance, while decreasing $\zeta$ places more emphasis on keeping the queues small (i.e., satisfying constraints). 

\texttt{DPP} is an \emph{online} algorithm. At each time $t$, it solves the simple Min-Weight problem in~\eqref{eq:dpp} and takes an action. A proof of queue stability under the \texttt{DPP} algorithm can be found in~\cite{luo2025semantic}. Notably, this approach achieves near-optimal performance with a low time complexity $\mathcal{O}(|\mathcal{U}|)$, where $|\mathcal{U}| = M + 1$ and $M$ is the number of sources in a multi-source system~\cite{luo2025semantic}. This makes it a promising method in large-scale systems.

\begin{figure*}[t!]
    \centering
    \begin{subfigure}{0.49\linewidth}
        \centering
        \includegraphics[width=0.95\linewidth]{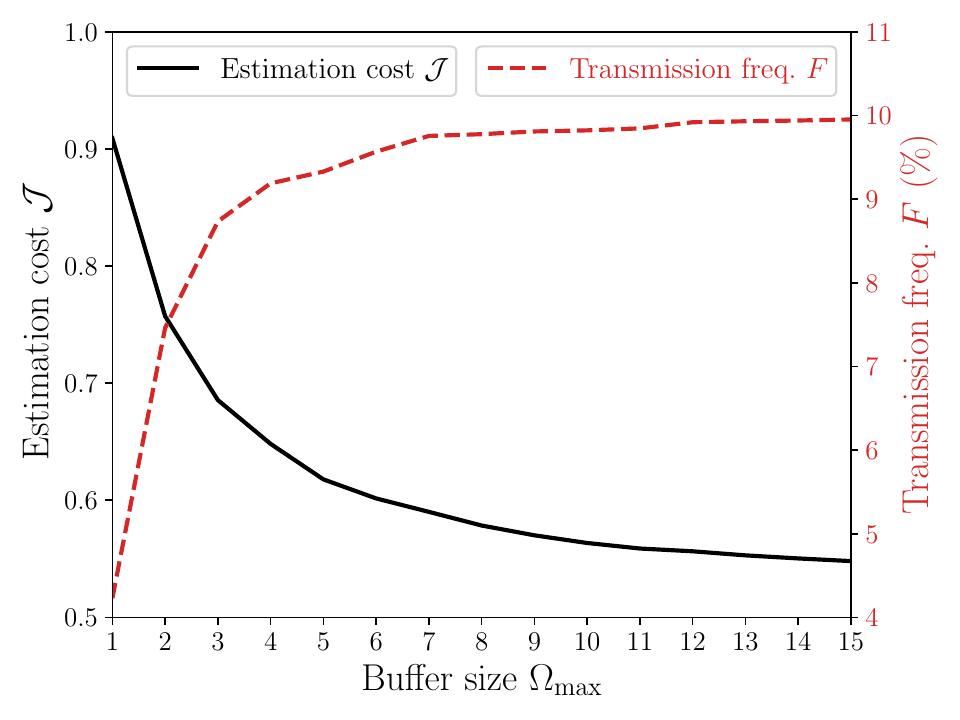}
        \caption{Token-based approach} 
        \label{fig:token-approach}
    \end{subfigure}
    \begin{subfigure}{0.49\linewidth}
        \centering
        \includegraphics[width=0.95\linewidth]{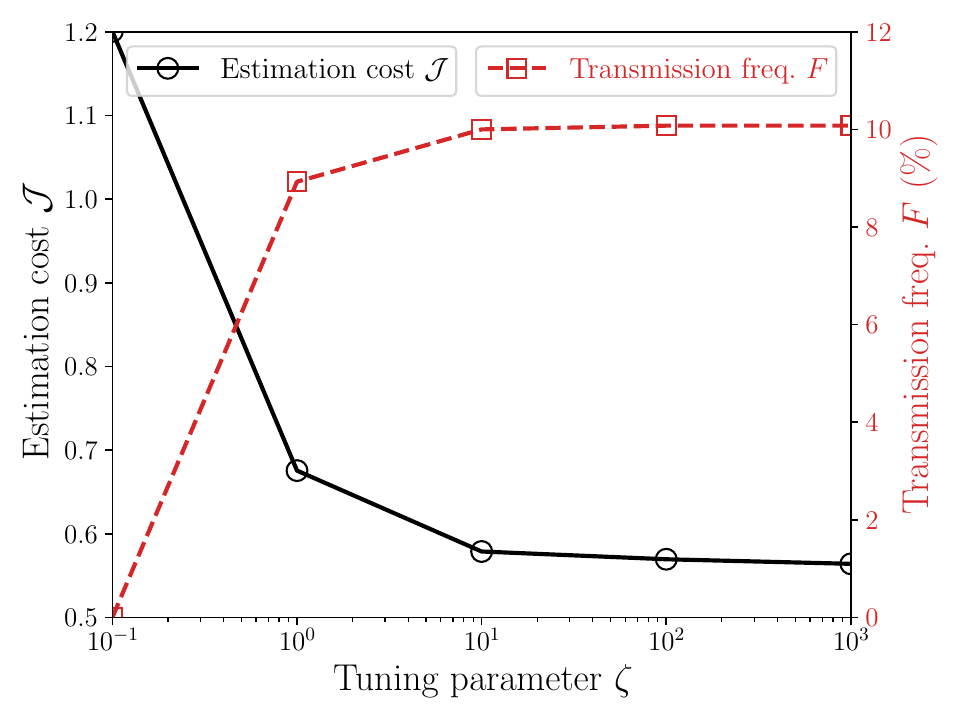}
        \caption{Lyapunov approach}
        \label{fig:lyapunov-approach}
    \end{subfigure}
    \caption{Performance of the token-based and Lyapunov approaches.}
    \label{fig:search-algorithm}
\end{figure*}

\subsubsection{Token-Based Relaxation}
This approach relaxes the constraint by introducing an auxiliary token variable $\Omega_t$, defined as \cite{stamatakis2023optimizing, delfani2025optimizing}
\begin{align}
    \Omega_{t+1} = \Big[\big[\Omega_t - U_t \big]_{0}^{\infty} + \Xi_t\Big]_0^{\Omega_{\max}}.\label{eq:token}
\end{align}
Specifically, the system earns one token, i.e., $\Xi_t = 1$, w.p. $F_{\max}$ at each time slot $t$. The tokens are stored in a finite buffer of size $\Omega_{\max}$ for future transmissions. If the buffer is full, new tokens are discarded. We assume that each transmission consumes one token, and the sensor must remain silent (i.e., $U_t = 0$) when the token buffer is empty (i.e., $\Omega_t = 0$). The virtual queue representation of the token process~\eqref{eq:token} is depicted in~Fig.~\ref{fig:virtual-queue-token}, where $\{\Xi_t\}_{t\geq 1}$ and $\{U_t\}_{t\geq 1}$ denote the arrival process and the controllable service process, respectively. 

Then, the constrained Problem~\eqref{problem:constrained-semantics-aware problem} is transformed into an unconstrained MDP characterized by the tuple $(\hat{\mathcal{S}}, \mathcal{U}, \hat{P}, c)$. The cost function $c$ and action space $\mathcal{U}$ are the same as the original constrained problem. The state space is defined as 
\begin{align}
    \hat{\mathcal{S}} = \{(S_t, \Omega_t): S_t\in\mathcal{S}, 0\leq \Omega_t \leq \Omega_{\max}\},
\end{align}
and the transition kernel is $\hat{P}(S_{t+1}, \Omega_{t+1}|S_t, \Omega_t, U_t)$. This MDP can be solved using standard dynamic programming techniques. It can be shown that the optimal policy has a switching structure as well, where the transmission thresholds depend on the instantaneous estimation error, the AoCE, and the number of available tokens.

Unlike the Lyapunov approach, which enforces constraint satisfaction through the objective function, the token-based approach directly imposes the constraint on the state and action spaces of the MDP. Since there is no need to trigger transmission when the system is synced or when the token buffer is empty, the effective service rate is strictly less than $F_{\max}$. Therefore, the constraint in \eqref{problem:constrained-semantics-aware problem} is met w.p. $1$. 

\subsubsection{Numerical Example} We now compare the performance of these approaches. Consider the numerical example examined in Section~\ref{sec:unconstrained-formulation} and set a communication budget $F_{\max} = 10\%$. The Lagrangian approach yields an optimal policy that randomizes between two switching policies and achieves a minimum cost of $\mathcal{J}^* = 0.5312$ at a transmission frequency of $F^* = 10\%$. Observe from Fig.~\ref{fig:search-algorithm} that the token-based and Lyapunov methods achieve near-optimal performance when their parameters are appropriately tuned.

Fig.~\ref{fig:token-approach} shows the performance of the token-based approach as a function of the buffer size $\Omega_{\max}$. It can be observed that small buffer sizes lead to conservative policies. As the buffer size increases, the system tends to fully utilize available transmission opportunities and approaches the minimum average cost. However, the time complexity grows exponentially as the token size increases. Hence, there is a trade-off between optimality and complexity. 

Fig.~\ref{fig:lyapunov-approach} shows the performance of the Lyapunov approach. When testing system performance, we take the average cost over $10^5$ samples. It can be observed that when the $\zeta$ is small, the estimation cost is negligible compared to the queue backlog, thus forcing the system to transmit less (i.e., reduce the arrival rate) to maintain a light-load queue. In contrast, when $\zeta$ is large, the system prioritizes cost minimization. However, it does not provide structural insight into the optimal policy or its transmission schedule. In contrast, the Lagrangian and token-based methods produce switching-type policies that facilitate analysis and implementation.

Table~\ref{table:complexity} compares the performance of these approaches.

\begin{table}[ht!]
\centering
\caption{Comparison of Different Approaches}
\begin{tabular}{ c | c | c | c | c}
    \toprule
     & \textbf{Complexity}
     & \textbf{Scalability}
     & \textbf{Structure}
     & \textbf{Optimality} \\
    \midrule
    Lagrangian & high & low & switching & optimal
    \\
    Token-based & medium & medium & switching & near-optimal\\
    Lyapunov & low  & high & unknown & near-optimal\\
    \bottomrule
\end{tabular}
\label{table:complexity}
\end{table}

\section{Applications and Results} \label{sec:applications}
This section showcases the benefits of incorporating the proposed framework in some indicative applications.

\subsection{Wireless Networked Control Systems}
This section illustrates the nonlinear aging effects in NCSs. Consider the remote estimation system described in Section~\ref{sec:nonlinear-aging}, where the receiver's error covariance is a monotonically increasing function of the AoI, denoted by $f(\Delta_t)$. We study remote estimation over a \emph{wearing channel} whose quality deteriorates due to natural aging and usage~\cite{jiping2025TACsubmission}. Such degradation phenomena are often observed in wireless communication systems with low-power devices or those exposed to harsh environments, where frequent communication or operational stress depletes energy reserves, accelerates hardware aging, and further impairs link reliability. Similar phenomena appear in biological neural networks, where synaptic efficacy declines with age and repeated activation~\cite{abraham1996metaplasticity}, and in quantum channels, where entanglement and coherence decay due to environmental interactions~\cite{schlosshauer2019quantum}.

Let $U_k \in \mathcal{U}$ denote the sensor’s decision at time slot $k$, where $\mathcal{U} = \{0, 1, 2\}$, and
\begin{itemize}
    \item $U_k = 0$ denotes the idle action. The channel is not used in slot $k$, and its reliability degrades due to natural aging. 
    \item $U_k = 1$ denotes the transmit action. Each transmission incurs a certain amount of wear on the channel, leading to an additional reduction in its reliability.
    \item $U_k = 2$ denotes the renewal action. Maintenance or replacement is performed to restore the channel's quality; however, this action takes time to complete.
\end{itemize}

We assume that channel renewal occupies $\Delta_\text{R} > 1$ consecutive slots. During renewal, the sensor must remain idle; that is, decisions are made only at the completion of each action. For ease of analysis, we introduce the decision epoch $t = 0, 1, \ldots$, where the sojourn time between the $t$th and $(t+1)$th epoch is
\begin{align}
    \ell(U_t) = \begin{cases}
        1, &U_t \in\{0, 1\},\\
        \Delta_\text{R}, &U_t = 2.
    \end{cases}\notag
\end{align}
The \emph{Age of Channel} (AoC) is defined as
\begin{align}
    \Theta_{t+1} = \begin{cases}
        \Theta_t +1, &U_t = 0,\\
        \Theta_t +\Theta_\text{D}, &U_t = 1,\\
        1, &U_t = 2,\\
    \end{cases}\notag
\end{align}
where $\Theta_\text{D}>1$ represents the amount of wear incurred with each transmission. The AoC summarizes the aging effect through the history of all sensor actions. 

The information state of this problem is
\begin{align*}
    S_t = (\Theta_t, \Delta_t).
\end{align*}
We define the cost of taking an action $U_t$ in state $S_t$ as the lump sum received prior to decision epoch $t+1$. Given that the cost is accrued during the renewal period, we define
\begin{align}
    \tilde{c}(S_t, U_t) = \begin{cases}
        f(\Delta_t), &U_t = 0,\\
        f(\Delta_t) + E_\text{T}, &U_t = 1,\\
        \sum_{r=0}^{\Delta_\text{R}-1}f(\Delta_t + r) + E_\text{R}, &U_t = 2,
    \end{cases}\notag
\end{align}
where $E_\text{T}$ and $E_\text{R}$ are the resource utilization costs associated with the transmit and renewal actions, respectively.

\begin{figure}[t!]
    \centering
    \includegraphics[width=0.8\linewidth]{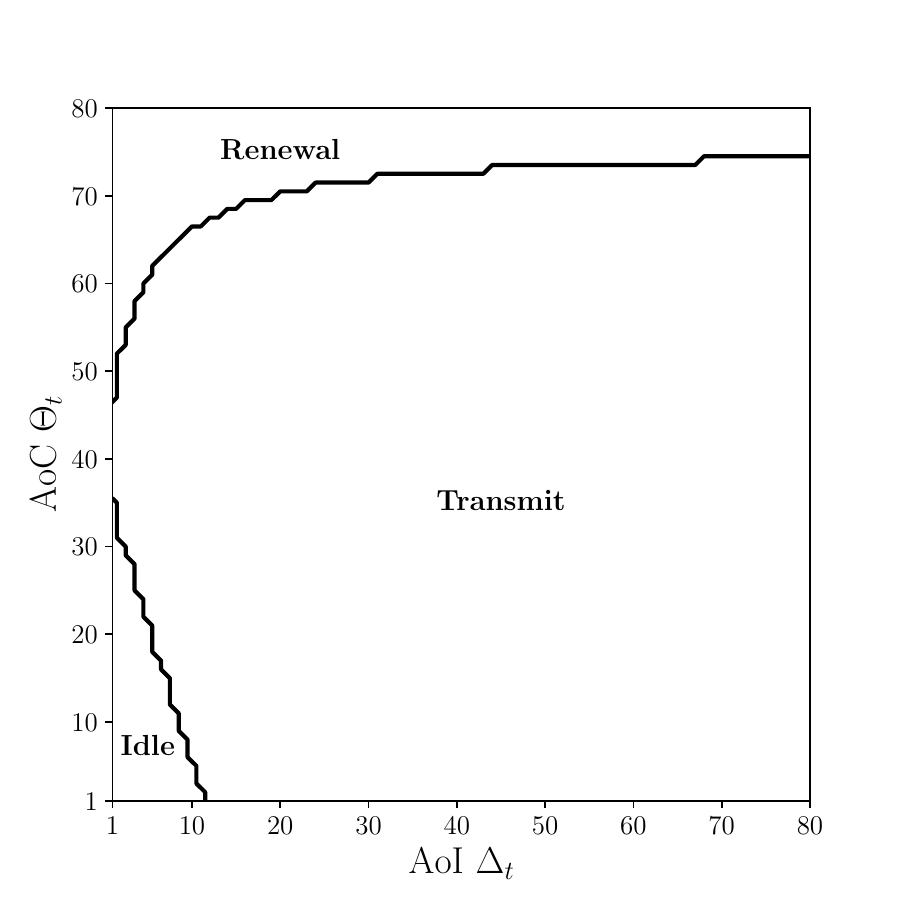}
    \caption{The optimal policy over an exponential wearing channel \cite{jiping2025TACsubmission}.}
    \label{fig:wearing-channel}
\end{figure}

The average estimation error is defined as
\begin{align}
    J(\pi) = \limsup_{T\rightarrow\infty}\frac{\mathbb{E}^\pi \Big[\sum_{t=0}^{T-1}\tilde{c}(S_t, U_t)\Big]}{\mathbb{E}^\pi \Big[\sum_{t=0}^{T-1}\ell( U_t)\Big]}.\notag
\end{align}
The goal is to find a policy $\pi^*$ that minimizes $\mathcal{J}(\pi)$. This is a semi-MDP problem, characterized by the tuple $(\mathcal{S}, \mathcal{U}, \tilde{P}, \tilde{c}, \ell)$, where $\tilde{P}$ is the transition kernel. A semi-MDP can be converted into an equivalent standard MDP through the uniformization method~\cite[Ch.~11.4]{puterman1994markov}. Let $\gamma$ be a scalar such that $0 < \gamma < \ell(u)/(1-\tilde{P}_{s,s}(u))$ for all $s\in\mathcal{S}$ and $u\in\mathcal{U}$ for which $\tilde{P}_{s,s}(u) < 1$. Define for all $s$ and $u$,
\begin{align*}
    c(s, u) = \frac{\tilde{c}(s,u)}{\ell(u)},\,P_{s,s^\prime}(u) = \begin{cases}
        \frac{\gamma\tilde{P}_{s,s^\prime}(u)}{\ell(u)}, &s\neq s^\prime,\\
        1-\frac{\gamma\big(1-\tilde{P}_{s,s}(u)\big)}{\ell(u)}, &s=s^\prime.
    \end{cases}
\end{align*} 
Then the semi-MDP $(\mathcal{S}, \mathcal{U}, \tilde{P}, \tilde{c}, \ell)$ is equivalent to the MDP $(\mathcal{S}, \mathcal{U}, P, c)$, which can be solved using the dynamic programming techniques developed in Section~\ref{sec:unconstrained-formulation}. It has been shown that the optimal policy is AoC-monotone~\cite{jiping2025TACsubmission}; that is, the optimal policy is weakly increasing (from idle to transmit to renewal) in AoC for any fixed AoI.

Fig.~\ref{fig:wearing-channel} shows the optimal policy results. It is observed that the optimal policy is AoC-monotone but not AoI-monotone. It aligns with the intuition that, when the channel condition is good and the information at the receiver is relatively fresh, the sensor can remain idle to save energy. In contrast, as the channel quality deteriorates and the information ages, the sensor takes more aggressive actions to improve estimation quality and channel reliability. 

\subsection{Gossiping Networks}
This section reports results on VAoI optimization in gossiping networks~\cite{delfani2024version}. Gossiping networks have become increasingly relevant for 6G and beyond systems, in which ultra-dense, ad-hoc, and resource-constrained deployments require lightweight and decentralized mechanisms for information dissemination~\cite{shah2009gossip}.

\begin{figure}[t!]
    \centering
    \includegraphics[width=0.7\linewidth]{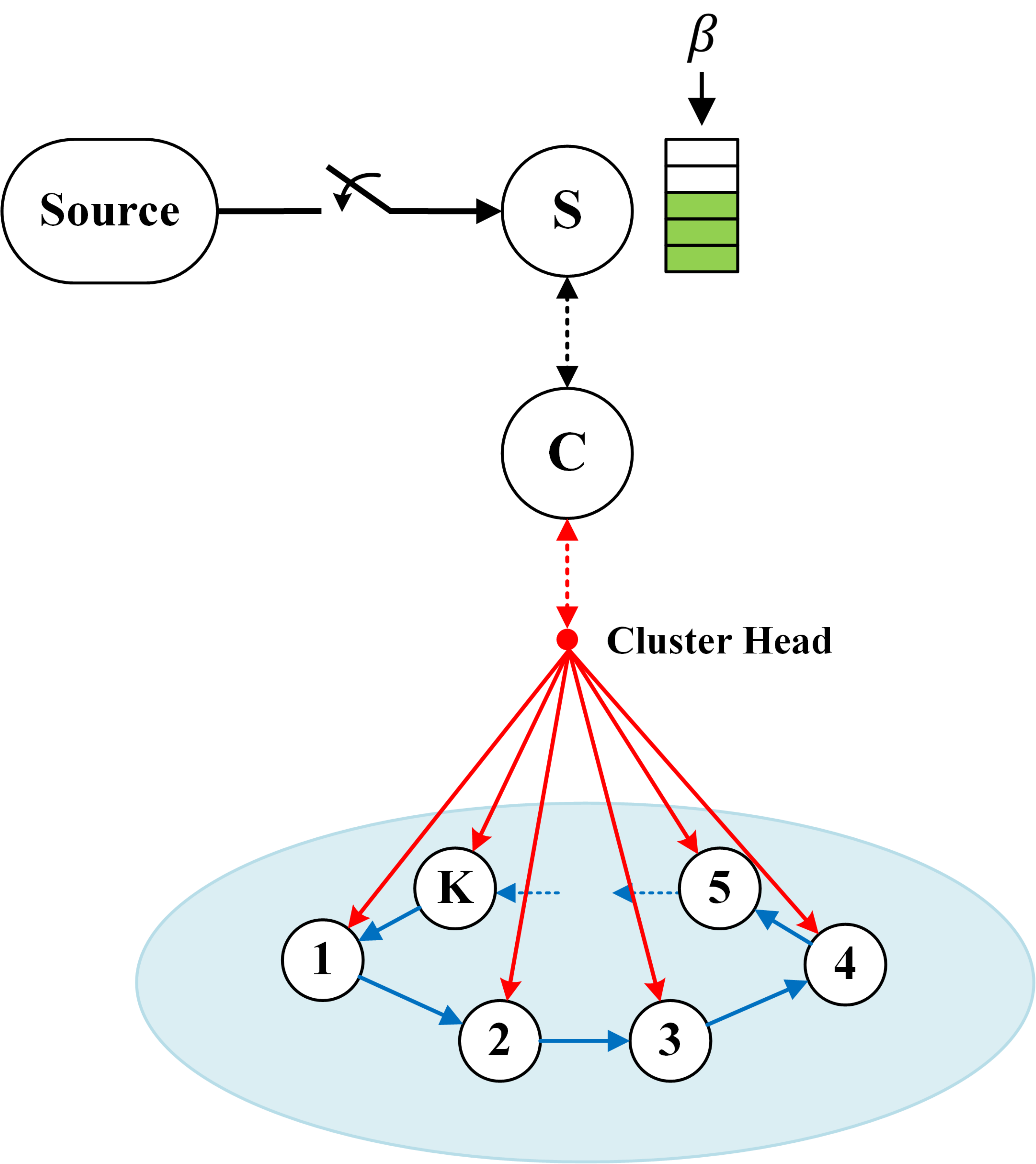}
    \caption{Semantics-aware status updates in a gossip network \cite{delfani2024version}.}
    \label{fig:GossipNet}
\end{figure}

We consider the system depicted in Fig.~\ref{fig:GossipNet}, comprising an energy-harvesting sensor (S), an aggregator (C), a cluster-head, and $K$ destination nodes. The version generation at the source follows a Bernoulli process $\{V^\text{S}_t\}$ at rate $p_g$, where $V^\text{S}_t = 1$ if a new version is generated at the source, and $V^\text{S}_t = 0$ otherwise. The sensor S is equipped with a rechargeable battery of size $B_{\max}$ and harvests energy from the environment. Energy arrivals follow a Bernoulli process $\{E_t\}$ with mean $\beta$. Each transmission consumes one unit of energy. Generating and transmitting a fresh update consumes one unit of energy. 

The aggregator C, located near the sensor, decides in each time slot whether to request a fresh update from S. Let $U_t$ denote C's action at time $t$: 
\begin{itemize}
    \item $U_t=1$ indicates that C requests a fresh update from S; if the sensor's battery is non-empty, S generates and sends a fresh status update, which is then forwarded by C to the requesting destination node;
    \item $U_t=0$ indicates that C serves the external request using its cached update.
\end{itemize}
Hence, the battery state evolves as
\begin{align*}
    B_{t+1} = \Big[\big[B_t - U_t\big]_{0}^{\infty} + E_t\Big]_{0}^{B_{\max}},
\end{align*}
where the operation $\big[x\big]^{b}_{a} = \min\big\{\max\{x, a\},b\big\}$ bounds a variable $x$ between an interval $[a, b]$. If the battery is empty, S cannot generate a fresh update; therefore, even if $U_t=1$, C must serve the request using the cached update. We assume that at most one request from the destination network is served in each slot.

Destination nodes generate requests through the cluster head, which collects these requests and instructs C to update, at most, one node per slot. The request process follows a categorical distribution: node $k$ requests service with probability $q_k$, and no request occurs with probability $1 - \sum_{k=1}^K q_k$.

The destination network uses a uni-directional ring topology for gossiping. Node $k$ receives an update from node $k-1$ with probability $\lambda_k$ in each slot, and node $1$ receives updates from node $K$. Gossiping is assumed to be error-free, and each transmission occupies one time slot. When a node receives an update directly from C, the outcome of the gossiping process in that slot does not affect its VAoI. This model captures the interplay among energy harvesting, caching, stochastic request arrivals, and gossip-based information dissemination, providing a realistic framework for analyzing VAoI-optimal policies in decentralized networks.

\begin{figure}[t!]
    \centering
    \includegraphics[width=0.9\linewidth]{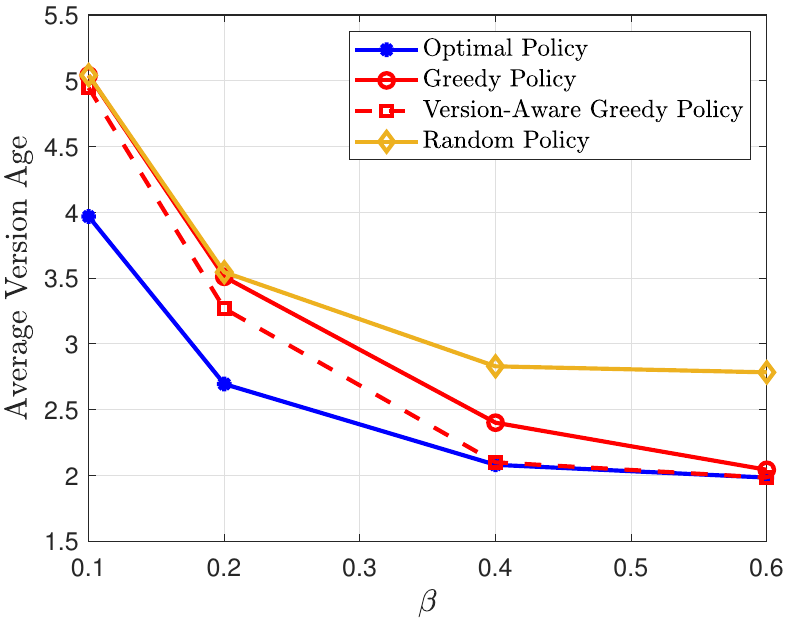}
    \caption{The average VAoI for a gossiping network with $K=3$ nodes, battery size $B_{\max}=5$, version update rate $p_g=0.5$, request service rate $q_1=0.1$, $q_2=0.2$, $q_3=0.3$, and gossiping rate $\lambda_{i=1,2,3}=0.2$~\cite{delfani2024version}.}
    \label{VersionAge_beta}
\end{figure} 

\begin{figure*}[t!]
    \centering
    \includegraphics[width=0.98\linewidth]{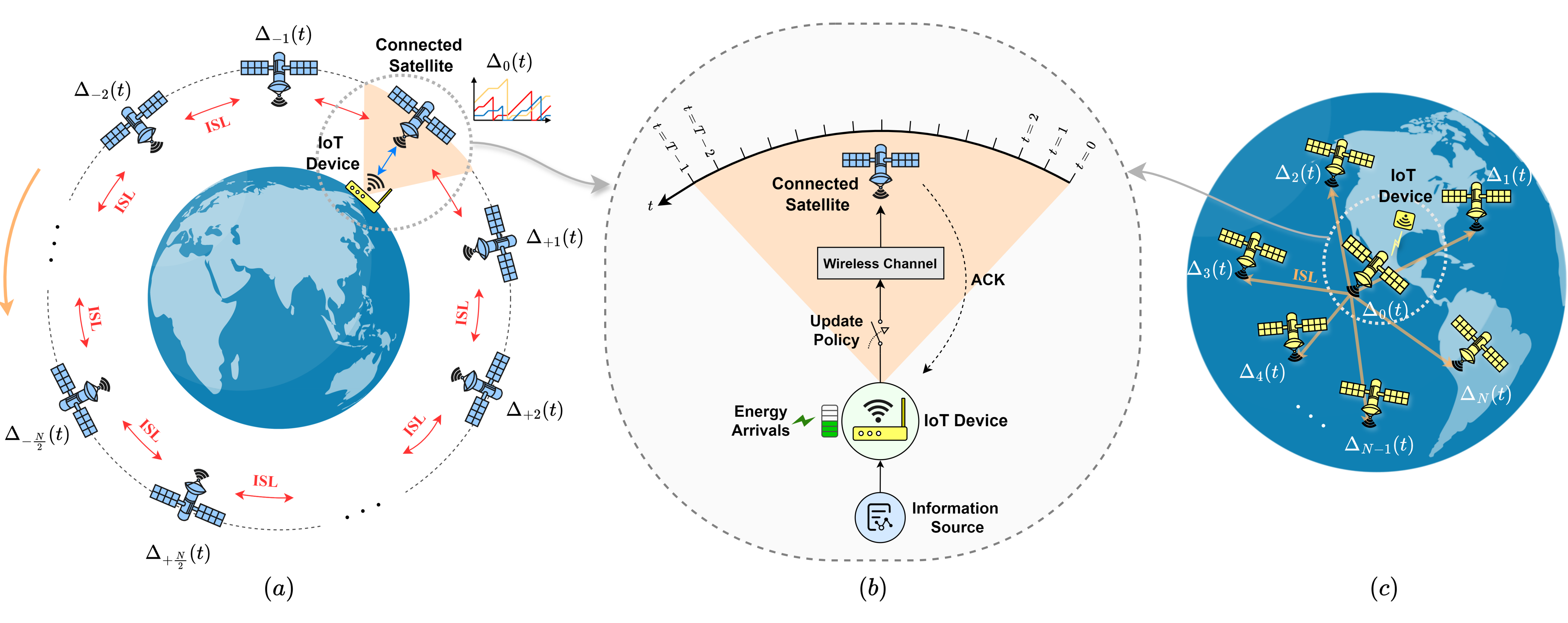}
    \caption{Status-update paths from an IoT device to an $(N+1)$-satellite LEO network: (a) ring, (c) star topology, and (b) shows the direct link from the device to the connected satellite~\cite{Delfani2025LEO}.}
    \label{fig_SysModelSat}
\end{figure*}

The VAoI at node $k$ evolves as
\begin{align}
    \Delta^k_{t+1}= V^\text{S}_t +
    \begin{cases}
        0, &\text{\small fresh update via C,} \\
        \Delta^\text{C}_t, &\text{\small cached update via C,} \\
        \min \left\{\Delta^k_t,\Delta^{k-1}_t \right\}, &\text{\small gossiping,}\\
        \Delta^k_t, & \text{\small not updated.}
    \end{cases} \notag
\end{align}
The average VAoI in the destination network is defined as
\begin{align*}
    \bar{\Delta}(\pi) = \limsup_{T\rightarrow\infty}\mathbb{E}^{\pi} \Bigg[
    \frac{1}{TK}\sum_{t=0}^{T-1}\sum_{k = 1}^K \Delta_t^k
    \Bigg],
\end{align*}
where $\pi = (U_0, U_1, \ldots)$ is the aggregator C's policy. The goal is to find an optimal policy to minimize the average VAoI. The information state for this MDP is
\begin{align*}
    S_t = (B_t, \Delta_t^1, \ldots, \Delta_t^K, \Delta_t^\text{C}).
\end{align*}
This MDP can be solved using the dynamic programming techniques developed in Section~\ref{sec:unconstrained-formulation}. Further details can be found in~\cite{delfani2024version}. We now summarize some key findings below.

(1) The optimal policy $\pi^*$ has a threshold structure: For each battery level $B_t$, there exists a threshold on the VAoI at the aggregator C, $\Delta_t^\text{C}$, such that the optimal action depends only on $(B_t, \Delta^\text{C}_t)$, and is independent of the VAoI at the destination nodes $\Delta_t^k$, $k = 1, \ldots, K$.

(2) The performance of the optimal policy is compared with three baseline policies: greedy, version-aware greedy, and random. In the greedy policy, the aggregator $C$ requests a fresh update whenever an external request arrives, regardless of the VAoI. The version-aware greedy policy is similar, but $C$ requests a fresh update only if the VAoI is non-zero. In the random policy, $C$ requests an update with a probability of $0.5$. As illustrated in Fig.~\ref{VersionAge_beta}, the VAoI-optimal policy consistently achieves the lowest average VAoI among all policies. When the energy harvesting rate $\beta$ is low, energy management becomes critical. In this regime, the optimal policy significantly outperforms the other policies by conserving energy and using it at more beneficial future time slots, whereas the baseline policies expend energy prematurely. When $\beta$ is high, the optimal policy acts more aggressively, behaving similarly to the version-aware greedy policy and achieving comparable VAoI performance.

\subsection{Satellite Communications}
Non-terrestrial networks (NTNs), such as LEO satellite constellations, represent a prominent class of systems expected to benefit significantly from semantic-aware communication~\cite{ETHER, delfani2025semantics, Delfani2025LEO}. These rapidly expanding constellations form dense, interconnected infrastructures capable of directly serving ground devices or acting as relays for global data exchange, caching, processing, and control. Yet, their stringent hardware, software, and energy limitations make efficient resource allocation crucial, especially under heavy data loads. This need becomes even more critical in time-sensitive cyber-physical and remote IoT scenarios, where timely transmission of informative rather than merely frequent data is essential for sustaining network performance.

To showcase the benefits, consider a system in which an energy-harvesting device transmits status updates to a network of $N+1$ LEO satellites, depicted in Fig.~\ref{fig_SysModelSat}. New source versions are generated according to a Bernoulli process with probability $p_g$. During each visibility window, the device connects to a Connected Satellite (CS) and decides, in each time slot, whether to transmit a fresh update ($U_t = 1$) or remain idle ($U_t = 0$) to conserve harvested energy. Energy arrivals follow a Bernoulli process with mean $\beta$. Arriving energy is stored in a battery whose state is denoted by $B_t$, with capacity $B_{\max}$. Each transmission consumes one energy unit.

We consider both \emph{ring} and \emph{star} LEO network topologies. In the ring topology, bidirectional Inter-Satellite Links (ISLs) enable deterministic, error-free propagation of updates along a ring of satellite nodes. In the star topology, the CS multicasts updates to $N$ one-hop neighbors via unreliable ISLs, each with success probability $\rho_n$. In both topologies, nodes store only the most recent update they receive.

The objective is to design an update policy $\pi$ to minimize the average VAoI in the LEO network. The average VAoI in the network is defined as
\begin{align}
    \bar{\Delta}(\pi) =\limsup_{T\rightarrow\infty}\mathbb{E}^{\pi} \Bigg[
    \frac{1}{(N+1)T}\sum_{t=0}^{T-1}\sum_{n\in\mathcal{N}} \Delta^n_t
    \Bigg],\notag
\end{align}
where $\Delta^n_t$ denotes the VAoI at the $n$th satellite at time $t$, and $\mathcal{N}$ is either $\mathcal{N}_\text{R} = \left\{-\frac{N}{2},-\frac{N}{2}\!+\!1,\cdots,\frac{N}{2}\!-\!1,\frac{N}{2}\right\}$ or $\mathcal{N}_\text{S} = \left\{0,1,2,\cdots,N\right\}$ for the ring and star topologies, respectively. This average VAoI can be expressed as~\cite{Delfani2025LEO}
\begin{align*}
    \text{Ring}:~\bar{\Delta}_\text{R}(\pi) &= \frac{N(N\!+\!2)}{4(N\!+\!1)}p_g + \bar{\Delta}_{0}(\pi),\\
    \text{Star}:~\bar{\Delta}_\text{S}(\pi) &= \frac{\sum_{n=1}^{N} \frac{1}{\rho_n}}{N+1}p_g + \bar{\Delta}_{0}(\pi), 
\end{align*}
where $\bar{\Delta}_{0}(\pi)$ is the average VAoI at the CS. Therefore, optimizing the average VAoI in the network for both topologies reduces to optimizing the average VAoI at the CS. The information state for this MDP is
\begin{align*}
    S_t = (B_t, \Delta_t^0),
\end{align*}
where $\Delta_t^0$ is the VAoI at the CS. The VAoI-optimal policy has a threshold structure: for each battery level $B_t$, the device triggers transmission only when the VAoI $\Delta_t^0$ exceeds a fixed threshold. 

The performance of the optimal policies is compared with two baselines: (i) the greedy policy, which transmits an update whenever energy is available, and (ii) the randomized stationary (RS) policy, which transmits with probability $\alpha$ in each time slot, provided the battery is not empty.

\begin{figure}[t!]
    \centering
    \includegraphics[width=0.95\linewidth]{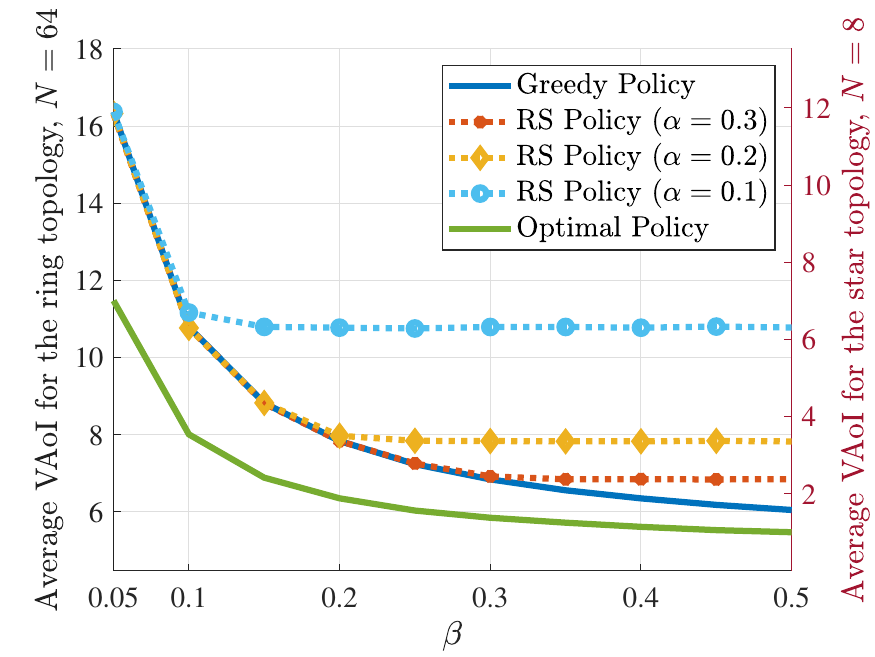}
      \caption{Average VAoI vs. $\beta$; same curves with different $y$-axes (ring: left, star: right), with $p_g = 0.3$, $B_{\max} = 20$, and $\rho_n = 0.7$ for all $n$~\cite{Delfani2025LEO}.}
      \label{fig_VAoIvsBeta}
\end{figure}

Fig.~\ref{fig_VAoIvsBeta} illustrates the average VAoI for both ring and star topologies as a function of the energy arrival probability $\beta$. Because the average VAoI values for the two topologies differ by a constant shift, there is an offset on the left and right $y$-axes. A key observation is that to achieve a target average VAoI (e.g., $8$ for the ring topology), the optimal policy requires an energy arrival rate of only $0.1$, which is half of what the greedy policy needs $0.2$. This result demonstrates that semantics-aware update policies can reduce energy consumption by up to 50\%: by transmitting fewer, better-timed updates, the satellite dissemination load is reduced, improving energy efficiency and extending system lifetime.

\section{Future Directions}
Looking forward, there are several promising research directions for advancing semantics-aware goal-oriented communication. As heterogeneous multi-agent systems continue to expand in scale and complexity, the need for communication protocols that operate effectively across physical and virtual environments becomes increasingly evident. In such settings, agents must coordinate and act based on information whose relevance depends not only on local observations but also on shared objectives and interactions with other entities. Thus, communication must adapt to the nature of these interactions, prioritizing important information that affects collective behavior and supports safe and effective collaboration.

A very interesting and important direction in goal-oriented communication systems where the goals may be unknown, partially known, or varying. In these cases, agents must not only communicate to accomplish the task but also learn it through interactions and communication. This challenge requires protocols that can infer, exchange, and refine representations of goals while simultaneously pursuing them. 

Achieving this vision requires combining several research domains, including integrating information-theoretic structure into neural architectures, developing benchmarks and evaluation methodologies for goal-driven communication, and designing robust schemes capable of operating under semantic, contextual, and temporal uncertainty. Additionally, there is a need for analytical tools that can formally characterize and verify communication behaviors, ensuring that the learned protocols remain safe, interpretable, and reliable. This is crucial, especially in domains such as autonomous driving, robotics, and industrial automation.

\section{Concluding Remarks}
As networks evolve toward 6G and beyond, goal-oriented semantics-aware communication offers a promising direction for achieving more efficient, reliable, and intelligent information exchange. This article has consolidated the conceptual foundations, methodological developments, and analytical tools that shape this emerging paradigm. By unifying classical distortion-based formulations, freshness-centric metrics, and recent advances in semantic value measures, we have highlighted how communication systems can move beyond accuracy and latency toward transmitting only task-relevant information with contextual, temporal, and operational significance. The analytical tools reviewed establish a coherent foundation for designing communication architectures tightly integrated with sensing, inference, and control. This synthesis will help crystallize design principles, stimulate interdisciplinary research across information theory, control, networking, and learning, and guide the development of practical semantic communication systems capable of supporting the next generation of real-time, data-driven intelligent networks.

\bibliographystyle{IEEEtran}
\bibliography{ref}

\end{document}